\documentclass[aps,prb,twocolumn]{revtex4-2}
\usepackage{amsmath}
\usepackage{graphicx}
\usepackage{mathtools}
\usepackage[english]{babel}
\usepackage[utf8]{inputenc}
\usepackage{hyperref}
\usepackage{textcomp} 
\usepackage{physics}
\usepackage{float}
\usepackage{dsfont}
\usepackage{enumitem}
\usepackage{MnSymbol}
\usepackage{comment}
\usepackage{color}

\begin{document}

\title{Tomography of near-field radiative heat exchange between\\mesoscopic bodies immersed in a thermal bath}
	
\date{\today}
	
\author{Florian Herz}
\email{florian.herz@institutoptique.fr} 
\affiliation{Laboratoire Charles Fabry, UMR 8501, Institut d'Optique, CNRS, Université Paris-Saclay, 2 Avenue Augustin Fresnel, 91127 Palaiseau Cedex, France}

\author{Riccardo Messina}
\affiliation{Laboratoire Charles Fabry, UMR 8501, Institut d'Optique, CNRS, Université Paris-Saclay, 2 Avenue Augustin Fresnel, 91127 Palaiseau Cedex, France}
\author{Philippe Ben-Abdallah}
\affiliation{Laboratoire Charles Fabry, UMR 8501, Institut d'Optique, CNRS, Université Paris-Saclay, 2 Avenue Augustin Fresnel, 91127 Palaiseau Cedex, France}
	
\begin{abstract}
A tomographic study of near-field radiative heat exchanges between a mesoscopic object and a substrate immersed in a thermal bath is carried out within the theoretical framework of fluctuational electrodynamics. By using the discrete-dipole-approximation method, we compute the power density distribution for radiative exchanges and highlight the major role played by many-body interactions in these transfers. Additionally, we emphasize the close relationship between power distribution and eigenmodes within the solid paving the way to applications for hot-spot targeting at deep sub-wavelength scale by shape optimization.
\end{abstract}
\maketitle

\section{Introduction}

Concentrating the heat emitted by a hot object placed at close distance to a base material if of significant relevance for practical uses such as nano-photolithography~\cite{Srituravanich_EtAl_2004}, to engrave patterning nanometer-scale structures on material surfaces, heat-assisted magnetic recording~\cite{Challener_EtAl_2009, Stipe_EtAl_2010} for data storage in hard-disk writing technology, or to make local temperature measurements using the near-field scanning thermal microscopy technique~\cite{DeWilde_EtAl_2006, Kittel_EtAl_2008, Jones_Raschke_2012}. Indeed, at distances smaller than the thermal wavelength $\lambda_\text{th}=\hbar c/k_{\rm B} T$ to the source, in which $k_{\rm B}$ is Boltzmann’s constant, the exchanged heat flux can surpass the flux predicted by the famous Stefan-Boltzmann’s law (blackbody limit~\cite{Planck_1914}) by several orders of magnitude~\cite{Polder_VanHove_1971, Volotkin_Persson_2007, Joulain_EtAl_2005} due to the contribution of evanescent photons which are superimposed on the propagative ones. In all these applications, a small section of the solid substrate is heated due to a super-Planckian heat flux to elevate its temperature beyond its melting point or beyond the Curie temperature of magnetic materials to demagnetize them locally. Therefore, to understand the underlying physics which govern the interactions of the electromagnetic field radiated by thermal emitters with microscope tips is of crucial importance. Despite recent progress in modeling these objects~\cite{Nguyen_Merchiers_Chapuis_2017, Narayanaswamy_Shen_Chen_2008, McCauley_EtAl_2012, Edalatpour_Francoeur_2016, Herz_EtAl_2018, Herz_Biehs_2022_2, BA_2019}, challenges remain in fully understanding this physics. Among the open questions is the description of many-body mechanisms~\cite{BA_Biehs_Joulain_2011, Messina_EtAl_2013, Biehs_EtAl_2021} which drive interactions and heat exchanges between mesoscopic objects and the field radiated by surrounding thermal emitters.

In the present work, we investigate this problem in detail within the framework of the fluctuational-electrodynamics theory~\cite{Rytov_Kravtsov_Tatarskii_1989}. We first introduce a theoretical model to analyze the radiative heat exchanges among coupled dipoles, a substrate, and the surrounding environment. We explore how individual resonant modes, potential surface modes, and collective modes contribute to these exchanges. Subsequently, employing the discrete dipole approximation method for thermal emitters~\cite{BA_2019, Draine_1988}, we compute the distribution of the power density for radiative exchanges within a mesoscopic object immersed in a thermal environment and positioned above a substrate at various distances. Thus, we provide a complete tomographic image of exchanges between each part of the object and its surrounding environment. By examining these power exchanges and the internal power spectra across the different regions within the solid, we reveal the diverse mechanisms driving heat exchanges between a tip and a substrate. Beyond the practical problem of the near-field thermal microscopy, we highlight the major role played by $N$-body effects on the energy exchanges between a mesocopic object and its surrounding environment. Additionally, we highlight the close connection existing between the power distribution within a solid and its eigenmodes thereby paving the way towards shape optimization for hot spot targeting.

The article is structured as follows: In Sec.~\ref{ch:Th_fram} we introduce our discrete-dipole-approximation method which we apply to two coupled dipoles in Sec.~\ref{ch:2p} both, in vacuum (Sec.~\ref{ch:vac}) and above a substrate (Sec.~\ref{ch:sub}). Subsequently, in Sec.~\ref{ch:disk} we discuss the tomography of a 2D disk spatially (Sec.~\ref{ch:pow_dis}) and spectrally (Sec.~\ref{ch:spec}). In the spectral analysis we also investigate a ring configuration with and without substrate (Sec.~\ref{ch:spec_ring}), special particles on a ring (Sec.~\ref{ch:special}), and the spatial distribution for special frequencies (Sec.~\ref{ch:spec_all}). In Sec.~\ref{ch:conc} we end up with our conclusion.

\section{Theoretical framework} \label{ch:Th_fram}

The system we consider is made of a collection of $N$ spherical particles of radii $R_\beta$ ($\beta \in \{1, ..., N\}$) and positions $\mathbf{r}_\beta$ placed in proximity to a substrate occupying the region $z<0$. Moreover, the particles are exposed to an environmental field, which can be thought as coming from the external boundaries, placed at large distance $z>0$ from particles and substrate. In the context of the dipolar approximation, we are going to describe each particle as an electric dipolar moment $\mathbf{p}_\beta$. To calculate the power dissipated into each particle in this configuration, we first need to determine the electric field and the electric dipole moments each $\mathbf{r}_\beta$. The electric field $\mathbf{E}$ at any position $\mathbf{r}$ can be decomposed into a contribution of the environmental field $\mathbf{E}_\text{env}$ and the ones which are induced by dipole moments $\mathbf{p}_\beta$ at the position $\mathbf{r}_\beta$, yielding
\begin{align}
\mathbf{E}(\mathbf{r}) & = \mathbf{E}_\text{env}(\mathbf{r}) + \mu_0 \omega^2 \sum_{\beta = 1}^N \mathds{G}_{\rm E}(\mathbf{r}, \mathbf{r}_\beta) \mathbf{p}_\beta .
\label{eq:field}
\end{align}
Herein, $\mathds{G}_{\rm E}(\mathbf{r}, \mathbf{r}_\beta)$ describes the electric Green's function at observation point $\mathbf{r}$ caused by an electric dipole moment at source point $\mathbf{r}_\beta$, $\omega$ represents the angular frequency and $\mu_0$ the vacuum permeability. Note that since we only take into account polar materials like silicon carbide (SiC), we can safely neglect the magnetic response of the system as it would be necessary in the case of metals~\cite{Dong_Zhao_Liu_2017, Tomchuk_Grigorchuk_2006, Dedkov_Kyasov_2007, Chapuis_EtAl_2008_1, Chapuis_EtAl_2008_2}. The electric dipole moments $\mathbf{p}_\beta$ can be decomposed into a fluctuating part $\mathbf{p}_{\beta, {\rm fl}}$ and an induced one $\mathbf{p}_{\beta, {\rm ind}}$, the latter depending on the electric field in Eq.~\eqref{eq:field} at position $\mathbf{r}_\beta$, so that
\begin{align}
\mathbf{p}_{\beta} & = \mathbf{p}_{\beta, {\rm fl}} + \varepsilon_0 \alpha_{\beta} \mathbf{E}(\mathbf{r}_\beta) \notag \\
& = \mathbf{p}_{\beta, {\rm fl}} + \varepsilon_0 \alpha_{\beta} \mathbf{E}_\text{env}(\mathbf{r}_\beta) + k_0^2 \alpha_{\beta} \sum_{\gamma = 1}^{N} \mathds{G}_{\rm E}(\mathbf{r}_\beta, \mathbf{r}_\gamma) \mathbf{p}_\gamma.
\label{eq:dip_mo}
\end{align}
In the last expression, $\varepsilon_0$ denotes the vacuum permittivity, $k_0 = \omega / c$ the wave number in vacuum, $c$ the velocity of light in vacuum, 
\begin{align}
\alpha_{\beta} & = \frac{\alpha_{{\rm CM}, \beta}}{1 - \text{i} \frac{k_0^3}{6 \pi} \alpha_{{\rm CM}, \beta}} , \\
\alpha_{{\rm CM}, \beta} & = 4 \pi R_\beta^3 \frac{\varepsilon_\beta - 1}{\varepsilon_\beta + 2},
\end{align}
the dressed and the Clausius-Mosotti polarizabilities in the weak form of the coupled dipole moments~\cite{Messina_EtAl_2013, AE_GarciaMartin_Cuevas_2017, Lakhtakia_1992, Albaladejo_EtAl_2010, Carminati_EtAl_2006}, respectively and $\varepsilon_\beta$ the dielectric permittivity of particle $\beta$. In the following, we will use SiC for both particles and substrate. Its dielectric permittivity can be described by a Drude-Lorentz model~\cite{Palik_1985}
\begin{align}
\varepsilon_{\beta}(\omega) & = \varepsilon_\infty \frac{\omega_{\rm LO}^2 - \omega^2 - {\rm i} \omega \Gamma}{\omega_{\rm TO}^2 - \omega^2 - {\rm i} \omega \Gamma}
\end{align}
with the following parameters: $\varepsilon_\infty = 6.7$, $\omega_{\rm LO} = 1.827 \times 10^{14} \, {\rm rad}/{\rm s}$, $\omega_{\rm TO} = 1.495 \times 10^{14} \, {\rm rad}/{\rm s}$, and $\Gamma = 0.9 \times 10^{14} \, {\rm rad}/{\rm s}$. 

We highlight that the Green's function appearing in Eq.~\eqref{eq:dip_mo} and detailed in Appendix \ref{AppA} and $\ref{AppB}$ explicitly contains the case $\beta = \gamma$ due to the contribution of the substrate (index s), whereas this contribution is neglected in the vacuum contribution (index 0), so that we can write for $\beta,\gamma=1,2,\dots,N$,
\begin{equation}\begin{split}
\mathds{G}_{\rm E}(\mathbf{r}_\beta,& \mathbf{r}_\gamma)  = \mathds{G}_{{\rm E, s}}(\mathbf{r}_\beta, \mathbf{r}_\beta) \delta_{\beta \gamma}\\
&  + \Bigl[ \mathds{G}_{{\rm E}, 0}(\mathbf{r}_\beta, \mathbf{r}_\gamma) + \mathds{G}_{\rm E, s}(\mathbf{r}_\beta, \mathbf{r}_\gamma) \Bigr] \left( 1 - \delta_{\beta \gamma} \right). 
\end{split}\label{eq:Green}\end{equation}
Equation \eqref{eq:dip_mo} can also be interpreted as one line of the block matrix equation
\begin{align}
\mathbf{p} & = \mathbf{p}_\text{fl} + \varepsilon_0 \boldsymbol{\alpha} \mathbf{E}_\text{env} + k_0^2 \boldsymbol{\alpha} \mathds{G}_{\rm E} \mathbf{p}
\label{eq:dip_mo_block1}
\end{align}
in which we define the block vectors $\mathbf{p} = (\mathbf{p}_1, ..., \mathbf{p}_N)^T$ and $\mathbf{E} = (\mathbf{E} (\mathbf{r}_1), ..., \mathbf{E} (\mathbf{r}_N))^T$ containing the dipole moments and electric fields, respectively, of each particle and the block matrices $\boldsymbol{\alpha} = {\rm diag}(\alpha_1, ..., \alpha_N)$ and $\mathds{G}_{\rm E}$ for which each block matrix component obeys $\mathds{G}_{{\rm E}, \beta \gamma} = \mathds{G}_{\rm E}(\mathbf{r}_\beta, \mathbf{r}_\gamma)$. With that we can recast Eq.~\eqref{eq:dip_mo_block1} in the following way
\begin{align}
\mathbf{p} & = \mathds{T}^{-1} \mathbf{p}_\text{fl} + \varepsilon_0 \mathds{T}^{-1} \boldsymbol{\alpha}_{\text{CM}} \mathbf{E}_\text{env} 
\label{eq:dip_mo_block}
\end{align}
with 
\begin{align}
\mathds{T}_{\beta \gamma} & = \left[\mathds{1} - k_0^2 \alpha_{\beta} \mathds{G}_\text{E,s}(\mathbf{r}_\beta \mathbf{r}_\beta) \right] \delta_{\beta \gamma} \notag \\
& \quad - k_0^2 \alpha_{\beta} \mathds{G}_{\rm E}(\mathbf{r}_\beta, \mathbf{r}_\gamma) (1 - \delta_{\beta \gamma}) .
\end{align}
In the same block-matrix notation, we can now rewrite Eq.~\eqref{eq:field} for each dipole by inserting Eq.~\eqref{eq:dip_mo_block} as
\begin{align}
\mathbf{E} & = \left[ \mathds{1} + k_0^2 \mathds{G}_{\rm E} \mathds{T}^{-1} \boldsymbol{\alpha} \right] \mathbf{E}_\text{env} + \mu_0 \omega^2 \mathds{G}_{\rm E} \mathds{T}^{-1} \mathbf{p}_\text{fl} .
\end{align}
The power $P_\beta$ dissipated into a given dipole $\beta=1,2,\dots,N$ can be written as
\begin{align}
	P_\beta & = \frac{1}{\pi} \int_0^\infty \text{d} \omega \, \omega \, \mathcal{C}_{\rm{pE},\beta} \\
	& = P_{\beta,{\rm dip} \rightarrow {\rm dip}} + P_{\beta,{\rm back} \rightarrow {\rm dip}} + P_{\beta,{\rm sub} \rightarrow {\rm dip}},
	\label{eq:contr_pow}
\end{align}
and is proportional to the imaginary part of the coupled dipole-field correlation function
\begin{equation}
	\mathcal{C}_{\rm{pE},\beta} = \sum_{i=x,y,z}\text{Im} \llangle p_{\beta,i} E_i^{*}(\mathbf{r}_\beta) \rrangle.
	\end{equation}
Note that we introduced the Fourier transform $f(t) = 2 {\rm Re} \left[ \int_{0}^{\infty} \frac{{\rm d} \omega}{2 \pi} f(\omega) e^{-{\rm i} \omega t} \right]$. The three contributions in Eq.~\eqref{eq:contr_pow} correspond to the heat dissipated into each single dipole due to heat transfer solely between the particles themselves, heat dissipated into each dipole due to the background, and heat dissipated into each dipole in presence of the substrate, respectively. Following the method outlined in Ref.~\cite{Messina_EtAl_2013}, each contribution can be recast under the form
\begin{align}
P_{\zeta \rightarrow{\rm dip},\beta} & = \int_0^\infty \frac{\text{d} \omega}{2\pi} \, \hbar\omega \, \mathcal{T}_{\beta, {\zeta \rightarrow {\rm dip}}}
\label{eq:T_coef}
\end{align}
for $\zeta \in \{ {\rm dip, back, sub} \}$, with the transmission coefficients
\begin{widetext}
\begin{align}
\mathcal{T}_{\beta, {{\rm dip} \rightarrow {\rm dip}}} & = 4\sum_{\gamma = 1}^N \sum_{i,j = x,y,z} n_{\gamma \beta} \frac{\chi_\beta}{|\alpha_{\beta}|^2} \left[\mathds{T}^{-1}\right]_{\beta \gamma,ij} \chi_\gamma \left[\mathds{T}^{-1 \dagger}\right]_{\gamma \beta, ji} , 
\label{eq:dd} \\
\mathcal{T}_{\beta, {{\rm back} \rightarrow {\rm dip}}} & = 4\sum_{\gamma, \delta = 1}^N \sum_{i,j,l = x,y,z} n_{\text{b} \beta} \frac{\chi_\beta}{|\alpha_{\beta}|^2} \left[\mathds{T}^{-1}\right]_{\beta \gamma,ij} \alpha_{\gamma} \, \mathcal{C}_{{\rm b}, \gamma \delta, jl} \, \alpha^{*}_\delta \left[\mathds{T}^{-1 \dagger}\right]_{\delta \beta, li} , 
\label{eq:db} \\
\mathcal{T}_{\beta, {{\rm sub} \rightarrow {\rm dip}}} & = 4\sum_{\gamma, \delta = 1}^N \sum_{i,j,l = x,y,z} n_{\text{s} \beta} \frac{\chi_\beta}{|\alpha_{\beta}|^2} \left[\mathds{T}^{-1}\right]_{\beta \gamma,ij} \, \alpha_\gamma \, \mathcal{C}_{{\rm s}, \gamma \delta, jl} \, \alpha^{*}_\delta \left[\mathds{T}^{-1 \dagger}\right]_{\delta \beta, li},
\label{eq:ds}
\end{align}
\end{widetext}
and introducing the Bose-Einstein occupation probabilities
\begin{align}
n_{\beta \gamma} & = \frac{1}{e^{\frac{\hbar \omega}{k_{\rm B} T_\beta}} - 1} - \frac{1}{e^{\frac{\hbar \omega}{k_{\rm B} T_\gamma}} - 1},
\end{align}
with the reduced Planck's constant $\hbar$, the Boltzmann constant $k_{\rm B}$, the temperature $T_\gamma$ of particle $\gamma$, and the temperatures $T_\text{b}$ and $T_\text{s}$ of background and substrate, respectively. The remaining correlation functions $\mathcal{C}_{\text{b/s},\beta \gamma, ij} = \llangle E_{i, {\rm b/s}}(\mathbf{r}_\beta) E_{j, {\rm b/s}}^{*}(\mathbf{r}_\gamma) \rrangle$ are~\cite{Eckhardt_1984, Messina_Antezza_2011_2}
\begin{equation}\begin{split}
&\mathcal{C}_{{\rm b},\beta \gamma, ij}  = \int \frac{\text{d}^2 k_\perp}{8 \pi^2} e^{\text{i} \mathbf{k}_\perp (\mathbf{x}_\beta - \mathbf{x}_\gamma)} \Theta_{\rm pr} \frac{e^{-\text{i} k_z (z_\beta - z_\gamma)}}{k_z} \\
& \quad \times \sum_{n = s,p} \Bigl( a^{-}_{n,i} + r_p e^{2 \text{i} k_z z_\beta} a^{+}_{n,i} \Bigr)\Bigl( a^{-}_{n,j} + r_p^{*} e^{-2 \text{i} k_z z_\gamma} a^{+}_{n,j} \Bigr),
\label{eq:Cb}
\end{split}
\end{equation}
and
\begin{equation}\begin{split}
&\mathcal{C}_{{\rm s},\beta \gamma, ij}  = \int \frac{\text{d}^2 k_\perp}{8 \pi^2} e^{\text{i} \mathbf{k}_\perp (\mathbf{x}_\beta - \mathbf{x}_\gamma)}  \\
& \quad \times \sum_{n = s,p} \biggl[ \Theta_{\rm pr} \frac{e^{\text{i} k_z (z_\beta - z_\gamma)}}{k_z} a^{+}_{n,i} a^{+}_{n,j} (1 - |r_n|^2) \\
& \quad + 2 \Theta_{\rm ev} \frac{e^{- |k_z| (z_\beta + z_\gamma)}}{|k_z|} a^{+}_{n,i} a^{-}_{n,j} \text{Im}(r_n) \biggr],
\label{eq:Cs}
\end{split}
\end{equation}
where we have introduced the projectors on the propagative and evanescent sections of the spectrum
\begin{align}
\Theta_{\rm pr} & = \Theta(k_0 - k_\perp), \\
\Theta_{\rm ev} & = \Theta(k_\perp - k_0).
\end{align}
For the correlation function of the dipole moments we have used the fluctuation-dissipation theorem~\cite{Eckhardt_1984, Messina_EtAl_2013}
\begin{align}
\llangle p_{\beta, i, {\rm fl}} p_{\gamma, j, {\rm fl}}^{*} \rrangle & = \hbar \varepsilon_0 \left( 1 + 2 n_{\beta} \right) \chi_\beta \delta_{\beta \gamma} \delta_{ij}
\end{align}
with
\begin{align}
\chi_\beta & = {\rm Im} \left( \alpha_\beta \right) - \frac{k_0^3}{6 \pi} |\alpha_\beta|^2.
\end{align}

Note that this is a very general formula to describe the heat flux dissipated into a collection of particles immersed in a vacuum background and in the presence of a substrate. It also allows to distinguish between the heat fluxes dissipated into each dipole meaning that we are able to discuss the spatial distribution of heat dissipated into each single dipole. In the following we will consider 1D and 2D systems and restrict ourselves to the simpler configuration of identical dipoles in the thermal-equilibrium configuration $T_1=...=T_N=T_{\rm p}$, so that $P_{{\rm dip} \rightarrow {\rm dip}} = 0$ holds. At first, we will evaluate our formula in the simplest scenario of two particles: this will allow us to highlight some physical mechanisms that we will also encounter for more particles. Later, we will discuss in detail the case of a 2D disk.

\section{Two particles} \label{ch:2p}

\begin{figure}
		\centering
		\includegraphics[width=0.25\textwidth]{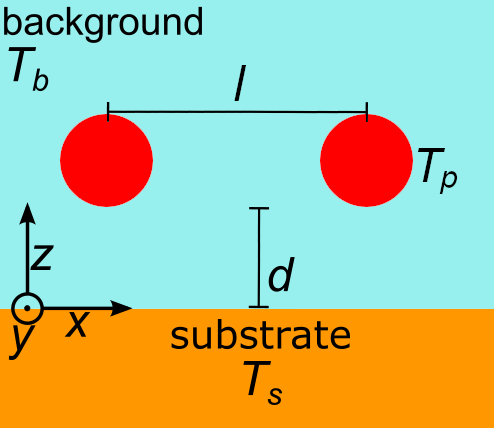}
		\caption{Sketch of two particles with temperature $T_{\rm p} = 298$ K and radius $R=19$ nm separated by a center-to-center distance $l = 3 R$ made of SiC at distance $d=200$ nm above a SiC substrate at temperature $T_{\rm s} = 323$ K immersed in a vacuum background at temperature $T_{\rm b} = 293$ K.}
		\label{fig:2p}
	\end{figure}

In order to validate our model and to get some first physical insight, let us start with the configuration of two particles in free space and above a substrate, shown in Fig.~\ref{fig:2p}. For the materials, we choose SiC for both the particles and the substrate, and we fix the temperatures at $T_{\rm p} = 298$\,K for both particles, $T_{\rm s} = 323$\,K for the substrate, and $T_{\rm b} = 293$\,K for the background. The particles are at edge-to-edge distance $d = 200$ nm from the substrate and $l=3R$ lateral distance to each other for an identical radius of $R = 19$\,nm, ensuring the validity of the dipole approximation. 

From a very fundamental point of view, this situation can be compared to two harmonic oscillators of identical mass $m$ coupled by a spring of stiffness $\kappa$, both bound to a wall with identical springs of stiffness $K$. This system is described by the system of differential equations
\begin{align}
m \ddot{x}_{1/2} & = - (K + \kappa) x_{1/2} + \kappa x_{2/1}
\end{align}
in which $x_{1/2}$ are the displacements of particle 1 and 2 and $\ddot{x}_{1/2}$ their second time derivatives. Rewriting this in matrix notation while making the ansatz $x_{1/2}(t) = c_{1/2} e^{{\rm i} \omega t}$ yields
\begin{align}
\begin{pmatrix} m \omega^2 - K - \kappa & \kappa \\ \kappa & m \omega^2 - K - \kappa \end{pmatrix} \begin{pmatrix} c_1 \\ c_2 \end{pmatrix} & = \begin{pmatrix} 0 \\ 0 \end{pmatrix} .
\end{align}
To fulfill this equation by setting the matrix determinant to zero, one obtains two resonance frequencies, namely $\omega_1 = \sqrt{\frac{K}{m}}$ and $\omega_2 = \sqrt{\frac{2 \kappa + K}{m}}$. The eigenvector $\mathbf{x}_1 = (1, 1)^t$ corresponding to $\omega_1$ describes both oscillators moving in the same direction and $\mathbf{x}_1 = (-1, 1)^t$ corresponding to $\omega_2$ describes both oscillators moving in opposite directions. While it is not straighforward to translate these quantities directly to the case of two nanoparticles described as dipoles, mainly because in the electromagnetic scenario we also have to take into account the different polarizations, we can conclude from this simple consideration that we should also find these two fundamental resonances in the spectrum of the two particles.

To show where we can find these resonances regarding our system of two dipoles, let us go back to Eqs.~\eqref{eq:field}-\eqref{eq:dip_mo} and recast them only considering the induced dipole moments, neglecting in other words $\mathbf{E}_{\rm env}$ and $\mathbf{p}_{\rm fl}$. In block matrix notation, one gets
\begin{align}
\boldsymbol{0} & = \mathds{M} \mathbf{p} , \\
\mathds{M} & = \mathds{1} - k_0^2 \alpha \mathds{G}_{\rm E}
\label{eq:M}
\end{align}
This equation demands ${\rm det}(\mathds{M}) = 0$ in the non-trivial case. The roots of this determinant will define the resonance frequencies of the system.

\subsection{In vacuum} \label{ch:vac}

We start by discussing the reference scenario of two particles in vacuum, i.e. in the absence of a substrate. In order to correctly describe the situation of an environmental field coming from both sides of the system ($z<0$ and $z>0$) we can take the limit in which the substrate is placed far away from the system of particles and it is replaced by an ideal blackbody. This amounts to neglect the evanescent contribution appearing in Eq.~\eqref{eq:Cs} and set $r_i = 0$. Under these assumptions the two correlation functions defined in Eqs.~\eqref{eq:Cb}-\eqref{eq:Cs} are identical, as well as the heat flux from both sides towards the particles ($P_{{\rm back} \rightarrow {\rm dip}}  = P_{{\rm sub} \rightarrow {\rm dip}} $). Then,  Eqs.~\eqref{eq:db}-\eqref{eq:ds} give the same result for $\beta=1,2$ and become
\begin{align}
\mathcal{T}_{\beta,{{\rm back/sub} \rightarrow {\rm dip}}} & = \pi k_0^2 n_{\text{bp}} \chi_{\rm p} \left( \mathcal{T}_\parallel + 2 \mathcal{T}_\perp \right),
\label{eq:2Pwos}
\end{align}
using the transmission coefficients
\begin{align}
\mathcal{T}_{\parallel/\perp} & = \frac{\bigl[ 1 + k_0^4 |\alpha|^2  |G_{{\rm E}, \parallel/\perp}|^2 \bigr] \frac{k_0}{6 \pi} + 2 k_0^2 {\rm Re} \bigl[ \alpha G_{{\rm E}, \parallel/\perp} \bigr] \mathcal{C}_{{\rm b}, 12, \parallel/\perp} }{|1 - k_0^4 \alpha^2 G_{{\rm E}, \parallel/\perp}^2|^2},
\end{align}
with
\begin{align}
G_{{\rm E}, \parallel} & = \frac{e^{{\rm i} k_0 l}}{2 \pi l} \frac{1 - {\rm i} k_0 l}{k_0^2 l^2}, \\
G_{{\rm E}, \perp}  & = \frac{e^{{\rm i} k_0 l}}{4 \pi l} \frac{k_0^2 l^2 + {\rm i} k_0 l - 1}{k_0^2 l^2}, 
\end{align}
and
\begin{align}
\mathcal{C}_{{\rm b},11/22, ij} & = \frac{k_0 }{6 \pi} \delta_{ij} , \\
\mathcal{C}_{{\rm b},12/21, \parallel} & = \frac{\sin(k_0 l) - k_0 l \cos(k_0 l)}{2 \pi k_0^2 l^3} , \\
\mathcal{C}_{{\rm b},12/21, \perp} & = \frac{k_0 l \cos(k_0 l) - (1 - (k_0 l)^2 \sin(k_0 l))}{4 \pi k_0^2 l^3} .
\end{align}
This already shows that due to the symmetry of the system the two transversal directions ($\perp$) contribute identically to the overall result while the longitudinal direction ($\parallel$) is different. 

\begin{figure}
		\centering
		\includegraphics[width=\columnwidth]{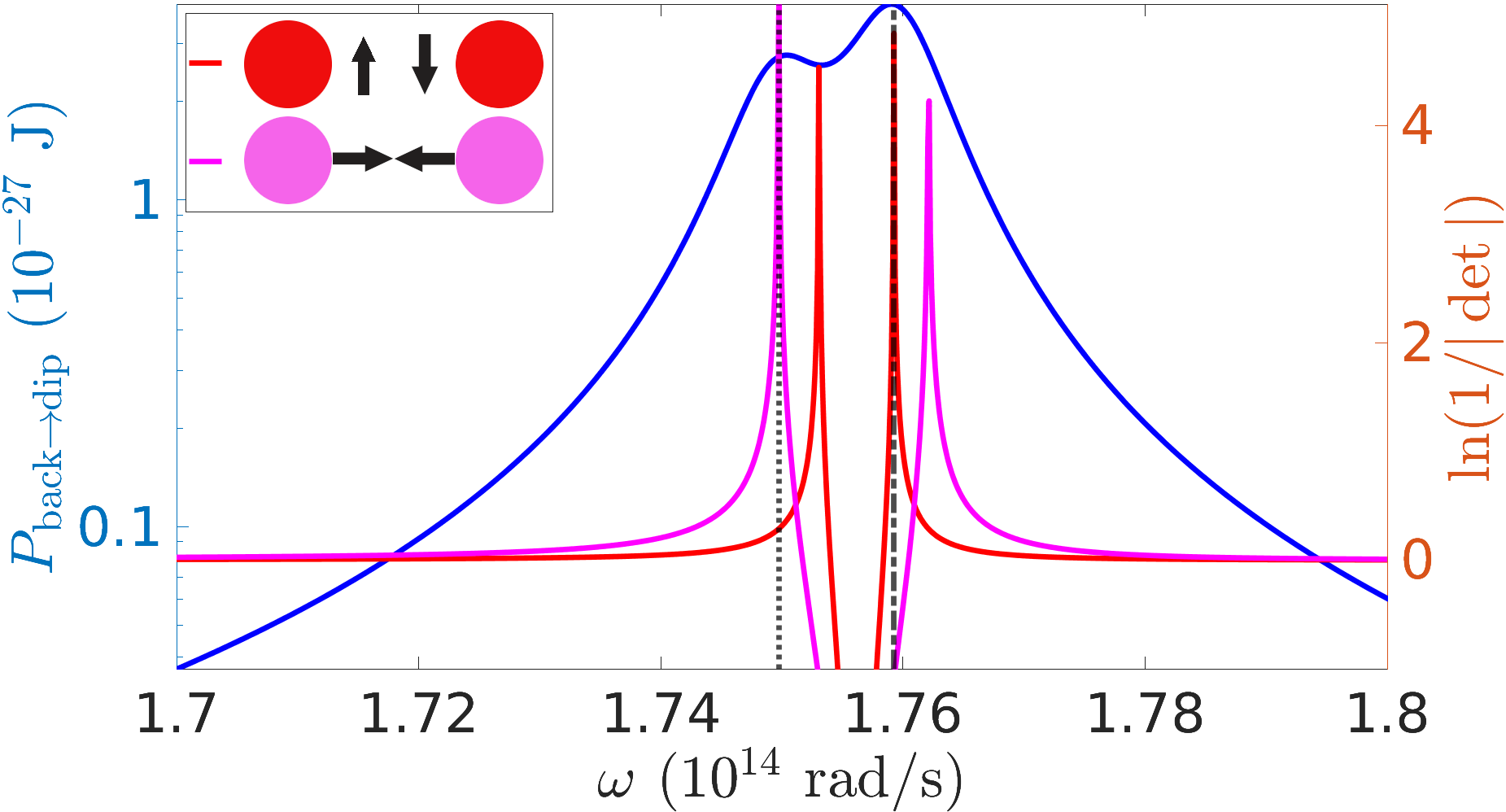}
		\caption{Spectral heat transfer between the two particles and the background (blue) shown with the determinant of matrices $\mathds{M}_{\perp/\parallel}$ (red/magenta) highlighting the resonance peaks for the transversal and longitudinal polarizations by the black dashed lines at $\omega_1 = 1.75 \times 10^{14} \, {\rm rad}/{\rm s}$ and $\omega_2 = 1.759 \times 10^{14} \, {\rm rad}/{\rm s}$.}
		\label{fig:2p_wos}
	\end{figure}

In Fig.~\ref{fig:2p_wos} we show the spectrum of the power absorbed by both dipoles according to Eq.~\eqref{eq:2Pwos}. There, we also depict the logarithm of the inverse of the determinant of the matrix defined in Eq.~\eqref{eq:M} but recast into 
\begin{align}
{\rm det} \left(\mathds{M} \right) & = {\rm det} \left(\mathds{M}_\perp \right)^2 {\rm det} \left(\mathds{M}_\parallel \right),
\end{align}
with 
\begin{align}
{\rm det} \left(\mathds{M}_{\perp/\parallel} \right) & = 1 - k_0^2 \alpha G_{{\rm E}, \perp/\parallel}.
\end{align}
Note that the polarizability $\alpha$ is now a scalar because we consider identical particles. Both factors $1 - k_0^2 \alpha G_{{\rm E}, \parallel/\perp}$ will exhibit two roots analogously to the model of two coupled harmonic oscillators. Obviously, the two transversal directions provide the same resonances which is mathematically described by the square. The resonances are shown in Fig.~\ref{fig:2p_wos} by the peaks of the inverse of the determinant. One can clearly see that there are four peaks, as expected, of which only two contribute to the overall spectrum, namely the lower resonance frequency of the longitudinal modes and the higher one of the transversal modes. That is because only these ``bright" modes contribute a dipole moment since they describe opposite movements as explained in the previous section~\cite{Herz_Biehs_2022_1, Downing_Mariani_Weick_2017}. The ``dark" modes describe translations of both dipoles in the same direction, not providing a dipole moment to the overall system and thus not coupling to the electromagnetic field.

\subsection{Above a substrate} \label{ch:sub}

In the presence of a substrate, the calculations become  more involved because the reflected part of the Green's function in Eq.~\eqref{eq:Gs} has to be considered as well. Additionally, the correlation functions also contain such a contribution taking the substrate into account. Both the reflected part of the Green's function and the two correlation functions are given for this configuration in Eqs.~\eqref{eq:Gs_2ws}-\eqref{eq:Dpm}. 

Doing the same analysis as in the previous section with respect to the determinant of matrix $\mathds{M}$ in Eq.~\eqref{eq:M}, but this time with respect to the $xz$ components of the reflected part of the Green's function, we do not have a longitudinal mode as before for two particles in vacuum but a coupling between the longitudinal mode and the transversal one in $x$ direction giving rise to a hybridized mode and yielding the following product of two determinants
\begin{align}
{\rm det} \left(\mathds{M} \right) & = {\rm det} \left(\mathds{M}_{\rm trans} \right) {\rm det} \left(\mathds{M}_{\rm mix} \right),
\end{align}
with 
\begin{align}
{\rm det} \left(\mathds{M}_{\rm trans} \right) & = \left(1 - k_0^2 \alpha G_{{\rm E,s},\perp}^{\rm id} \right)^2 - k_0^2 \alpha \left( G_{{\rm E,s},y}^{\rm dif} + G_{{\rm E},\perp} \right)^2,
\end{align}
and
\begin{align}
{\rm det} \left(\mathds{M}_{\rm mix} \right) & = \bigl[ \alpha^2  k_0^4 G_{\rm E,s,mix}^2 - \bigl( 1 - \alpha k_0^2 \bigl(G_{\rm E,s,x}^{\rm dif} + G_{{\rm E},\parallel} \bigr) \bigr) \notag \\
& \quad \times \bigl( 1 + \alpha k_0^2 \bigl( G_{\rm E,s,z}^{\rm dif} - G_{{\rm E},\perp} \bigr) \bigr) \bigr] \notag \\
& \quad \times \bigl[\alpha^2 k_0^4 G_{\rm E,s,mix}^2 - \bigl( 1 - \alpha k_0^2 \bigl( G_{\rm E,s,z}^{\rm dif} + G_{{\rm E},\perp} \bigr) \bigr) \notag \\
& \quad \times \bigl(1 + \alpha k_0^2 \bigl(G_{\rm E,s,x}^{\rm dif} - G_{{\rm E},\parallel} \bigr) \bigr) \bigr] .
\end{align}
Due to the coupling, the determinant of $\mathds{M}_{\rm mix}$ has now four roots.

\begin{figure}
		\centering
		\includegraphics[width=\columnwidth]{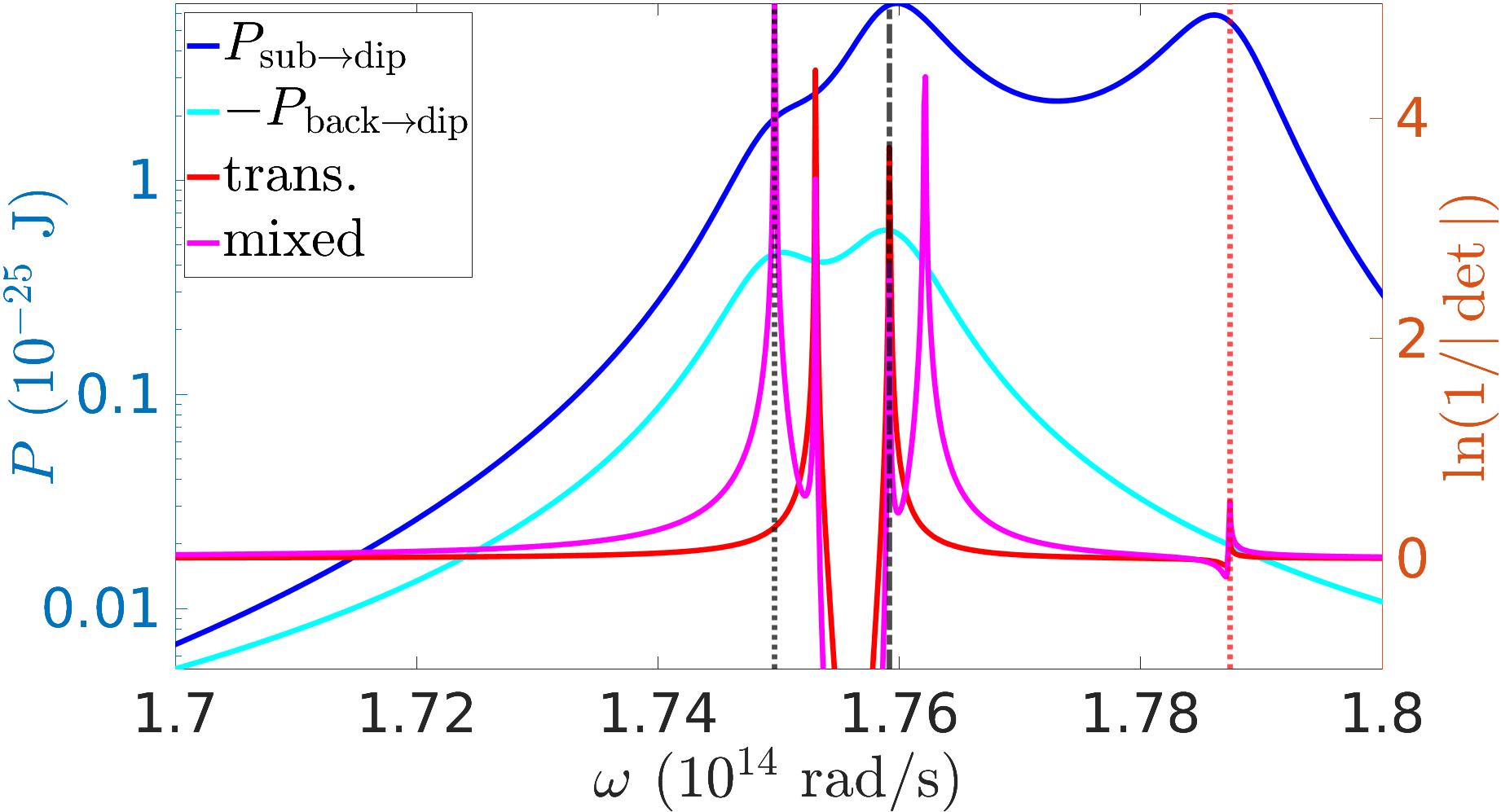}
		\caption{Spectral heat transfer between the two particles and the substrate (dark blue) or the background (light blue) shown with the determinant of matrices $\mathds{M}_{\perp/{\rm mix}}$ (red/magenta) highlighting the resonance peaks for the transversal and mixed polarizations by the black and red dashed lines at $\omega_1 = 1.75 \times 10^{14} \, {\rm rad}/{\rm s}$, $\omega_2 = 1.759 \times 10^{14} \, {\rm rad}/{\rm s}$, and $\omega_{\rm SPhP} = 1.787 \times 10^{14} \, {\rm rad}/{\rm s}$.}
		\label{fig:2p_ws}
	\end{figure}

The results for the heat exchange between the two particles and the background as well as with the substrate are shown in Fig.~\ref{fig:2p_ws}. Note that, due to our choice of temperatures, the spectral power between particles and background is negative. Therefore, we show $-P_{{\rm back} \rightarrow {\rm dip}}$ instead. The spectrum for the heat flux between the particles and the background (light blue) looks the same as in Fig.~\ref{fig:2p_wos} apart from an overall amplification. The spectral peaks for this contribution are at the same positions as for the case without substrate and can also be found in the heat flux between the particles and the substrate (dark blue). As shown by the inverse determinants, these resonance frequencies correspond to two roots of the determinant. The fact of not seeing all of them through a contribution to the spectral flux can be attributed to the existence of dark and bright modes as in the case without substrate. 

In addition to these two peaks for both spectral heat fluxes, there is also a third peak in the spectral heat flux between the dipoles and the substrate. It corresponds to the well known resonance frequency of the surface phonon polariton (SPhP) of SiC at $\omega_{\rm SPhP} = 1.787 \times 10^{14} \, {\rm rad}/{\rm s}$ at which the parallel polarized reflection coefficient in Eq.~\eqref{eq:rp} has a maximum due to $\varepsilon_{\rm sub} (\omega_{\rm SPhP}) \approx -1$. Since this mode characterizes a strong coupling between substrate and particles, it can only be found in the heat flux between the dipoles and the substrate and not in the one with the background. Note that both determinants also exhibit a root at $\omega_{\rm SPhP}$ due to the inclusion of the reflected part of the Green's function.

Translating these validations for two particles to the upcoming case of a 2D disk, there will be resonances stemming from the eigenmodes of the disk itself which coincide with the case without substrate. In the spectral heat flux between the particles and the substrate there will be a peak at $\omega_{\rm SPhP}$ which will dominate for the near field due to its evanescent character. 

\section{2D disk} \label{ch:disk}

\begin{figure}
		\centering
		\includegraphics[width=0.25\textwidth]{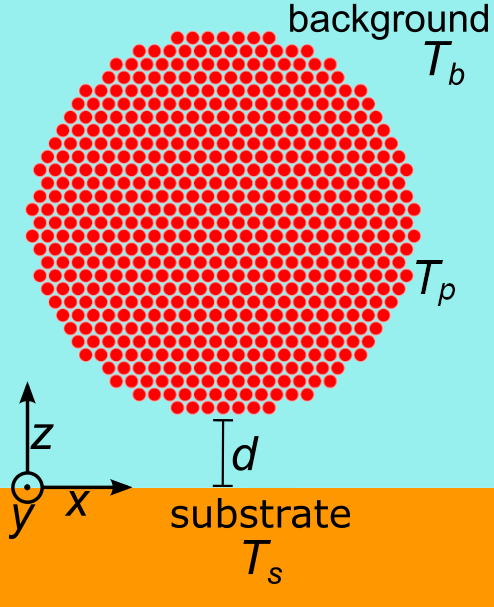}
		\caption{Sketch of a 2D disk with radius $R = 500$ nm at temperature $T_{\rm p} = 298$ K at distance $d=200$ nm above a SiC substrate at temperature $T_{\rm s} = 323$ K immersed in a vacuum background at temperature $T_{\rm b} = 293$ K. All $N = 583$ particles have a radius of $r =19$ nm and are made of SiC.}
		\label{fig:disc}
	\end{figure}

In the next step, we want to consider a 2D disk of radius $R=500$\,nm as depicted in Fig.~\ref{fig:disc}. The parameters for background and substrate remain the same as before. The particles are identical in temperature $T_{\rm p} = 298$ K, radius $r=19$\,nm, and both particles and substrate are made of SiC. In order to fill a disk of given radius $R$ as densely as possible with nanoparticles we consider a close-packing arrangement and keep only the nanoparticles which are fully inside the surface of the disk. Using this approach, a disk completely filled with particles corresponds to $N=583$ particles. The radius $r$ in with respect to radius $R$ is chosen such that $N r^2 / R^2 = 0.84$. For comparison, this ratio is $N r^2 / R^2 = 0.9$ for $r=3$\,nm and $N=24895$ so that the chosen radius of $r=19$\,nm already represents a disk whose surface area is almost completely filled by dipoles and, therefore, provides a reasonable approximation of the disk geometry in the spirit of DDA.

\begin{figure}
		\centering
		\includegraphics[width=\columnwidth]{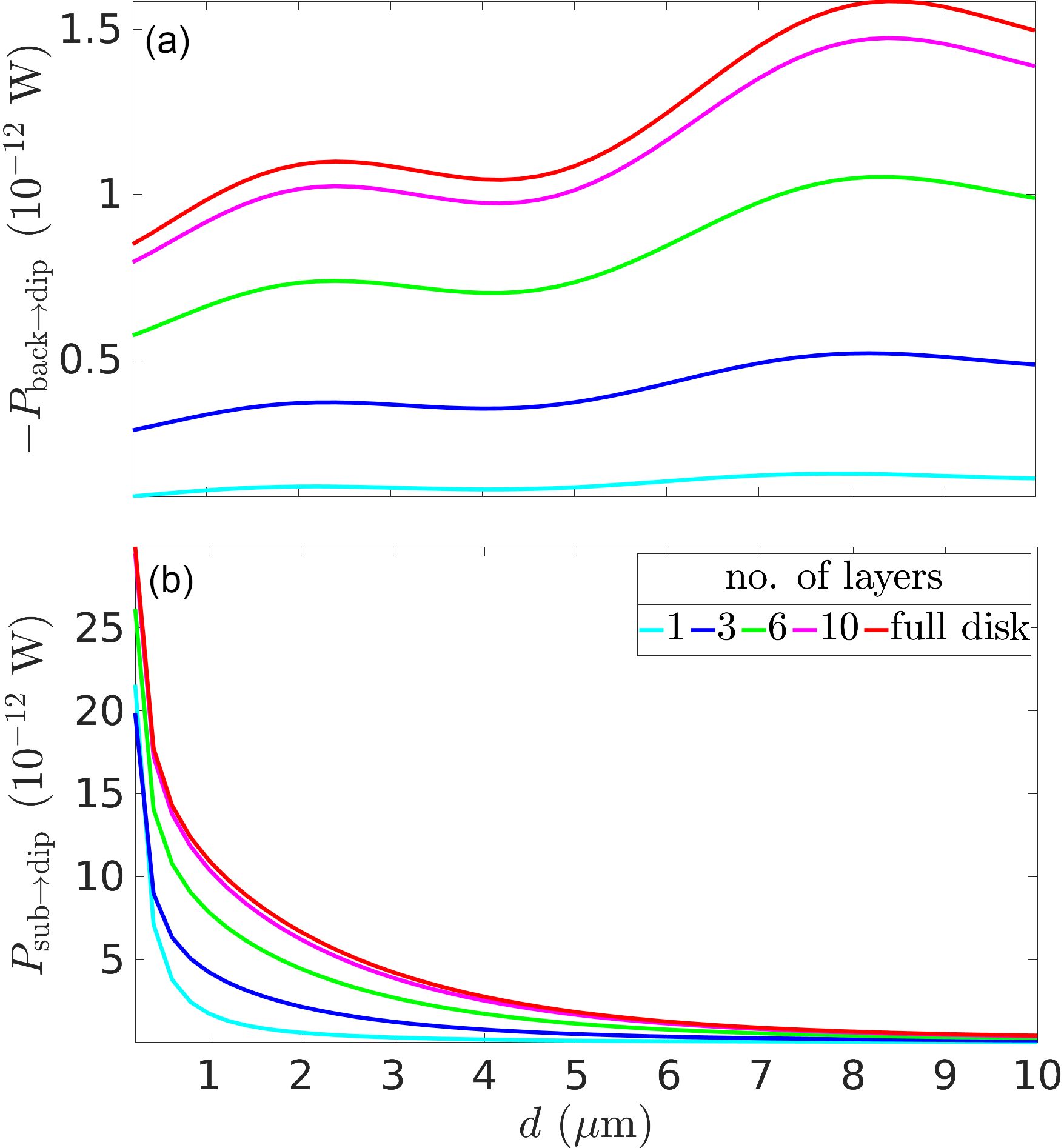}
		\caption{$-P_{{\rm back} \rightarrow {\rm dip}}$ (a) and $P_{{\rm sub} \rightarrow {\rm dip}}$ (b) for different numbers of layers starting with one ring on the outside and adding more towards the inside until we obtain a full disk.}
		\label{fig:layers}
	\end{figure}

In Fig.~\ref{fig:layers} we show on the total power
\begin{equation}
	P_{\zeta \rightarrow{\rm dip}} = \sum_{\beta=1}^NP_{\beta,\zeta \rightarrow {\rm dip}},
\end{equation}
for $\zeta \in \{ {\rm back, sub} \}$ absorbed by the system of $N$ dipoles, and show both contributions $-P_{{\rm back} \rightarrow {\rm dip}}$ and $P_{{\rm sub} \rightarrow {\rm dip}}$ while varying the distance between the disk and the substrate for different amounts of layers starting with only a ring of particles on the outside of the disk in Fig.~\ref{fig:disc} and adding more rings towards the inside until we end up with the full disk. One can clearly see that $P_{{\rm sub} \rightarrow {\rm dip}}$ dominates in the near field and up to distances $d < 6\,\mu$m. This distance also coincides with the coherence length of the SPhP mode of the SiC substrate which is the dominating coupling mechanism in the near field between substrate and particles. On the contrary, in the far-field regime $P_{{\rm back} \rightarrow {\rm dip}}$ becomes more important, resulting in a sign flip for the total received flux. Let us also stress that because of the non-monotonic behavior of $P_{{\rm back} \rightarrow {\rm dip}}$, due to interference between the radiation directly emitted away from substrate and particles and the one reflected at the substrate, this non-monotony will be imprinted on the overall result in the far field as a signature of this contribution because $P_{{\rm sub} \rightarrow {\rm dip}}$ is decaying exponentially. $P_{{\rm back} \rightarrow {\rm dip}}$ also increases for larger distances showing an attenuation effect due to coupling to the substrate. Note that this increase for $P_{{\rm back} \rightarrow {\rm dip}}$ in absolute values will asymptotically reach the blackbody limit for even larger distances. Additionally, we want to stress that although all curves share indeed a the same qualitative behavior, the ratio between them does not coincide with the ratio of the number of particles involved in each configuration.

From a numerical point of view it is also interesting to note that $P_{{\rm sub} \rightarrow {\rm dip}}$ converges faster towards the result of the full disk, which happens already for nine layers, than $P_{{\rm back} \rightarrow {\rm dip}}$ for which 11 layers are required, which corresponds to all particles but the seven at the center of the disk. Especially in the near field in which $P_{{\rm sub} \rightarrow {\rm dip}}$ dominates, one could reduce the computation time by taking less layers into account. Nevertheless, we will see later that this does not hold for a qualitative discussion because the different configurations will have strikingly different spatial power distributions.

\subsection{Spatial power distribution} \label{ch:pow_dis}

Now, we want to turn to the spatial distribution of the heat fluxes. Again, we will look at the evolution of a ring of particles towards the full disk to see whether certain particles or regions are more important than others. The parameters are as before. We consider three different distances between the (layered) disk and the substrate to cover the near field ($d=200$\,nm), the far field ($d=10\,\mu$m), and the intermediate regime ($d=1\,\mu$m). Here, we restrict ourselves to three figures for the setup of only one outer ring, a crown of four layers, and the full disk~\cite{Video}.
\begin{widetext}
	
\begin{figure}[H]
		\centering
		\includegraphics[width=0.7\textwidth]{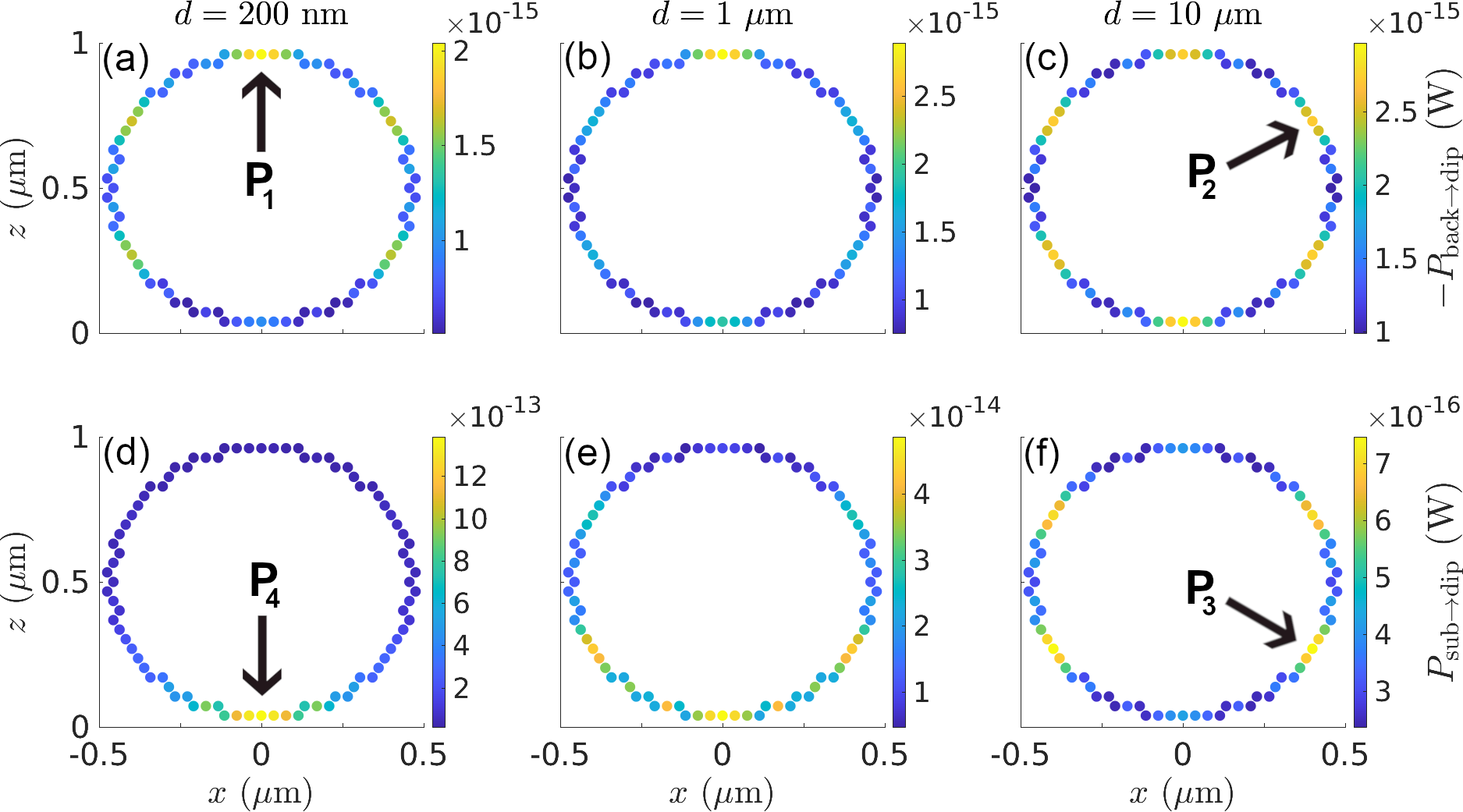}
		\caption{Spatial power distribution for one layer. (a)-(c): Heat flux between particles and background. (d)-(f): Heat flux between particles and substrate. The columns correspond to three different distances given above the upper panel. For better reference in chapter \ref{ch:special} we already highlight the particles of concern for later spectral analysis by P$_1$-P$_4$.}
		\label{fig:1l}
	\end{figure}
\begin{figure}[H]
		\centering
		\includegraphics[width=0.7\textwidth]{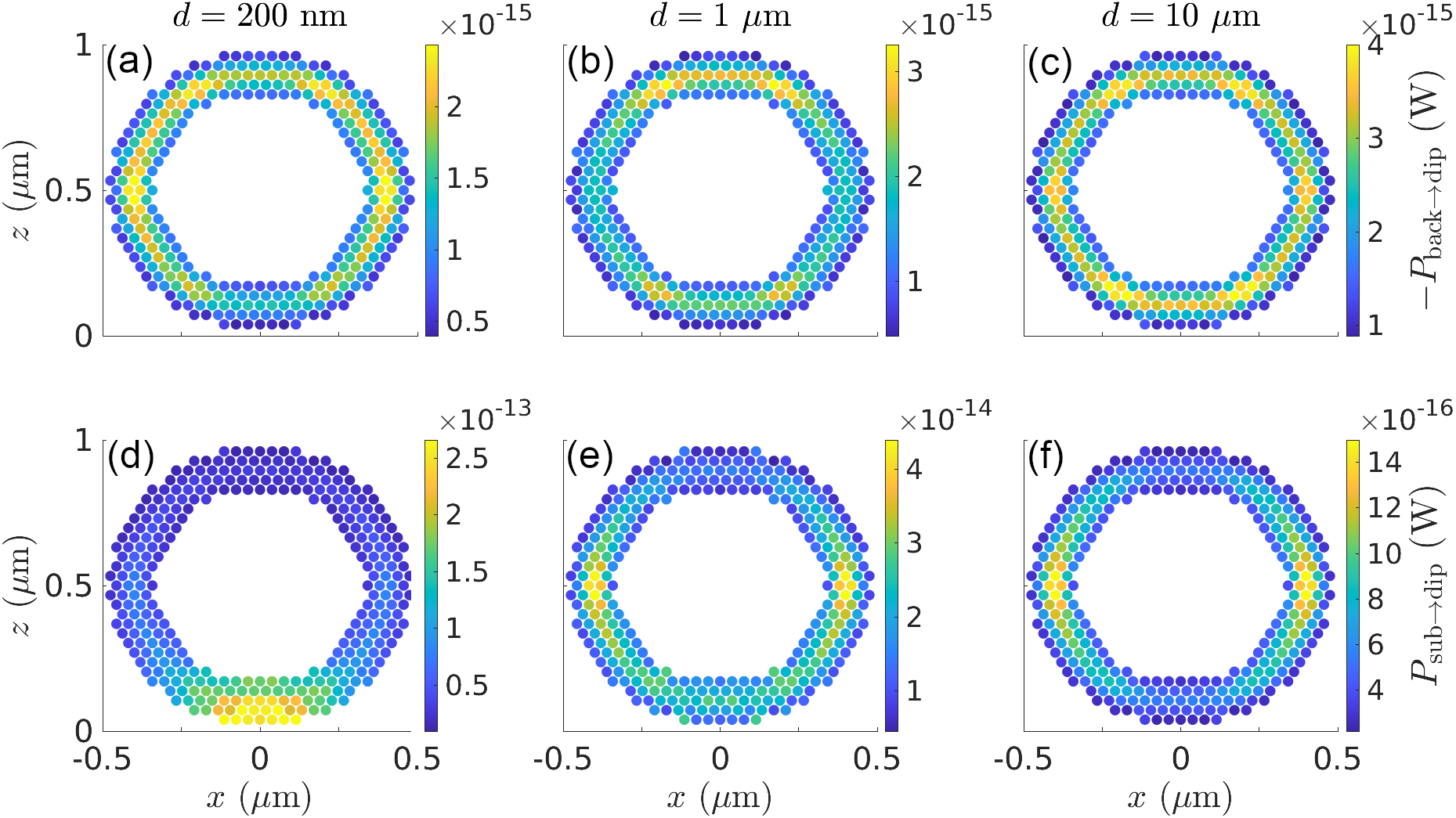}
		\caption{Spatial power distribution for four layers.}
		\label{fig:8l}
	\end{figure}
\begin{figure}[H]
		\centering
		\includegraphics[width=0.7\textwidth]{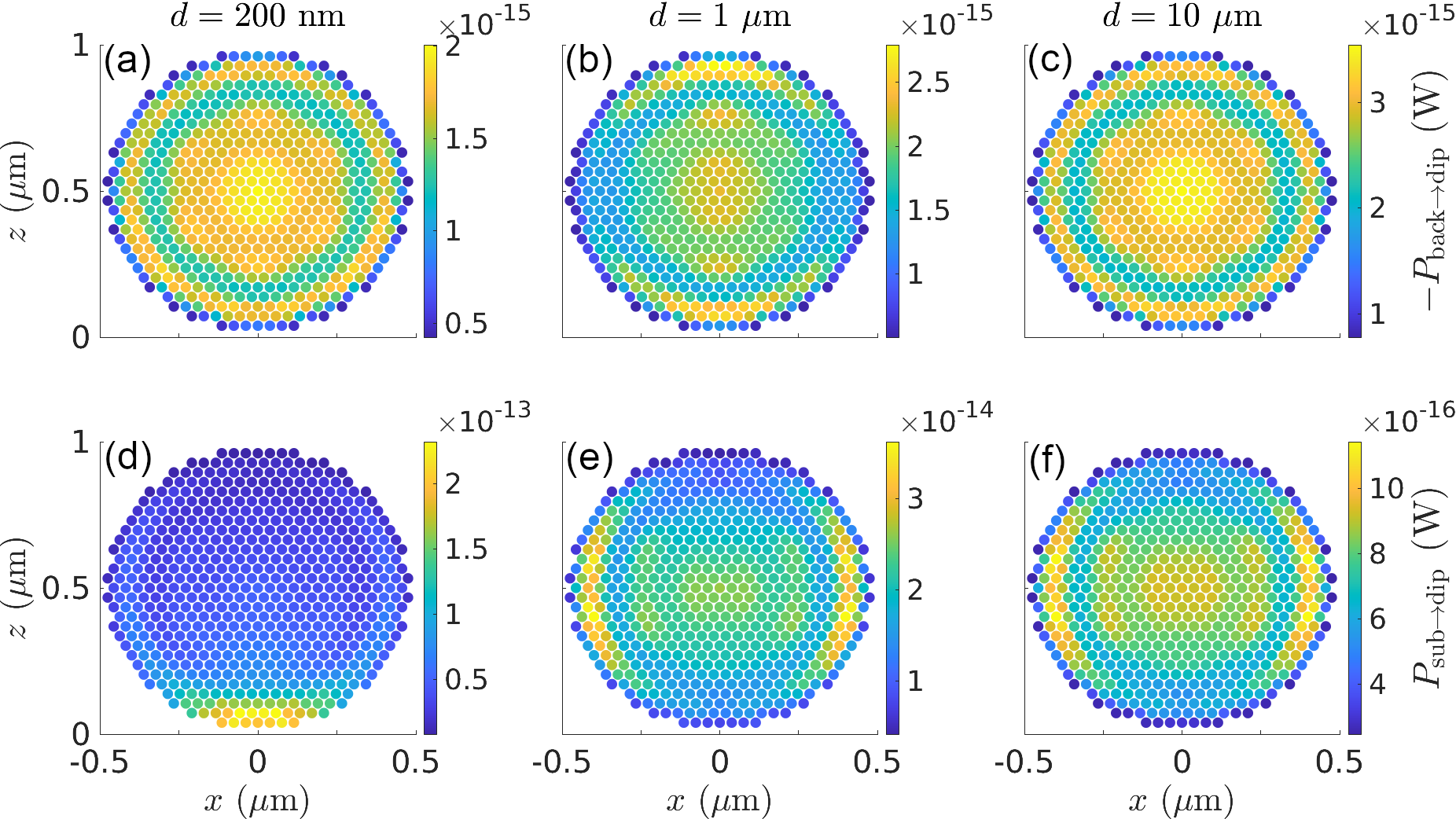}
		\caption{Spatial power distribution for the full disk.}
		\label{fig:full}
	\end{figure}
\end{widetext}

Let us start with the description of the lower panels (d)-(f) of Figs.~\ref{fig:1l}, \ref{fig:8l} and \ref{fig:full}, corresponding to the heat flux between particles and substrate. In all configurations one can clearly see for $d=200$\,nm that the particles close to substrate (bottom) experience the largest flux from the substrate. This is clearly a signature of the SPhP mode due to the strong coupling in this near-field regime. For larger distances $d$ this signature vanishes and we are left with bright spots in the middle. This is reminiscent of configurational modes of each disk layout in accordance to our findings for two particles in Sec.~\ref{ch:2p}, and suggests that here we are indeed observing an indirect evidence of eigenmodes of the ring, crown, and disk, respectively. Interestingly, we can also find bright spots at the bottom if we stay below four layers in the intermediate regime. The bright spot at the bottom, then, smears out towards the middle and center. This is clearly a many-body effect due to more particles that can couple with each other and distribute the incoming heat flux from the substrate for more particle layers. For less layers the closest particles still obtain more heat due to evanescent waves like frustrated modes in this intermediate regime. Due to poorer coupling, especially in the case of a single layer where there are only two nearest neighbors for each particle, the bottom part still obtains much more power than the upper part. Nevertheless, the eigenmode signature dominating in the far field can be already seen for configurations with less than four layers. This already represents a competition between the SPhP mode and the configurational eigenmodes of the system.

The upper panels (a)-(c) of Figs.~\ref{fig:1l}, \ref{fig:8l} and \ref{fig:full}, corresponding to the heat flux between particles and background, are also worth discussing because it shows counter-intuitive behavior. First of all, the upper part of the ring, crown, and disk emits more heat in the near-field regime for each configuration. For less layers this is more strictly bound to the very top and smears out again for more layers until we find brighter spots in the middle and center but still the upper part emits more power than the lower part. This shows that the substrate attenuates heat transfer towards the background due to coupling which is stronger for particles closer to the substrate. Interestingly, this brighter upper part is even more pronounced for the intermediate regime in all configurations. In the far field we expect again the eigenmodes to dominate the spectrum to whom we ascribe the spatial distribution of $P_{{\rm back} \rightarrow {\rm dip}}$ for each configuration. In the intermediate regime we would have expected something between near and far field meaning that the upper part is still more pronounced but also a stronger signature of the eigenmodes. This could, for example, hint towards the influence of frustrated modes. As another interesting feature, we find in the far field that the lower part for configurations with less than four layers is more pronounced than the upper region. 

We would like to stress the presence of the bright spots at the center which always occur apart from the cases of only a few layers and from $P_{{\rm sub} \rightarrow {\rm dip}}$ in the near field. This is, as we will show later in more detail, related to the eigenmodes of the configuration which is even an important feature if we only remove the central particle of the full disk. Additionally, it is important to stress the exception for configurations with only a small number of layers. For less than four layers there are many striking differences. To show the mechanisms behind this spatial power distribution, we will now analyze these configurations spectrally.

\subsection{Spectral analysis of the dissipated power} \label{ch:spec}

\begin{figure}[H]
		\centering
		\includegraphics[width=0.5\textwidth]{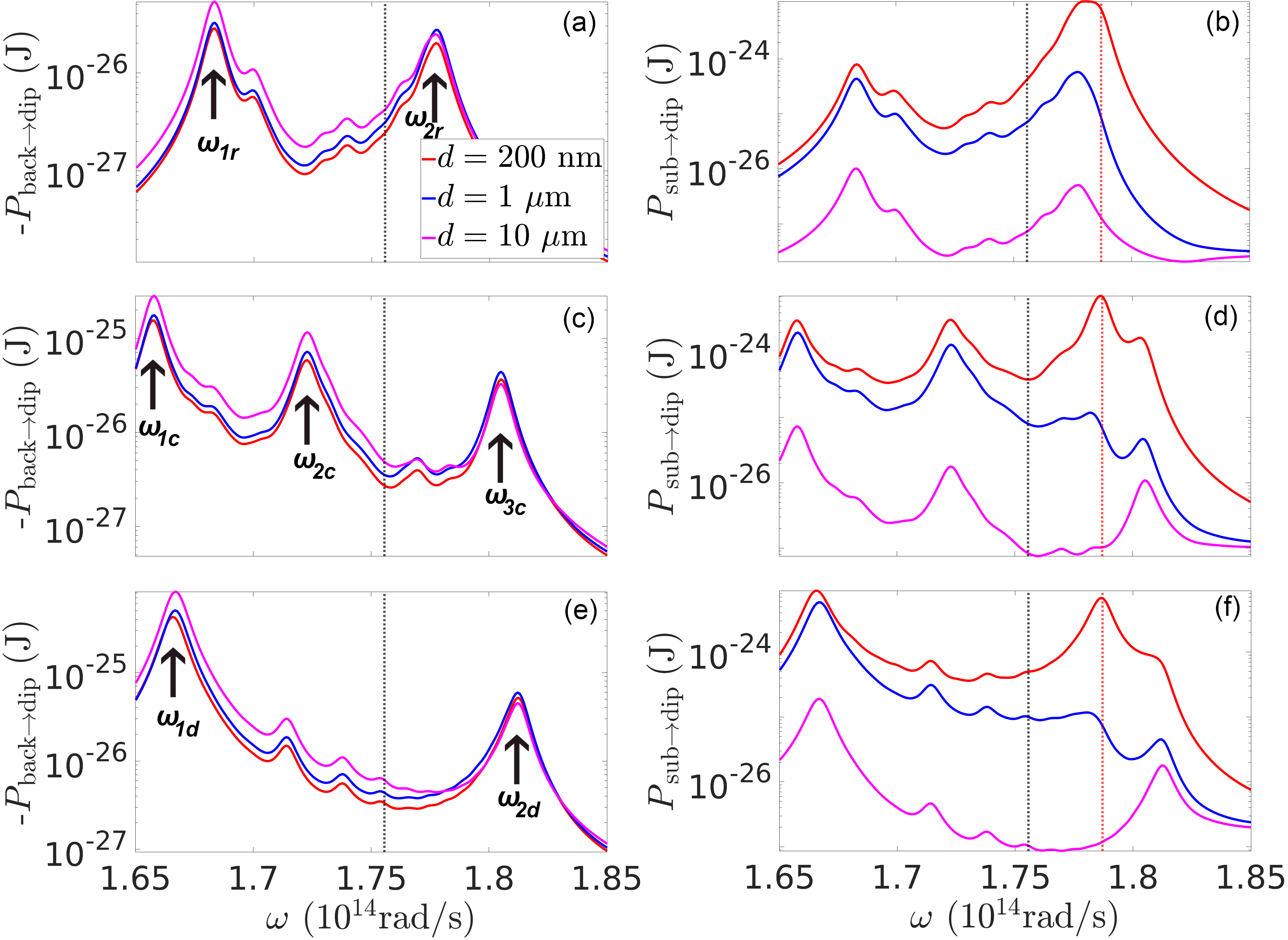}
		\caption{Spectral heat flux between the particles and the background (left, turned positive) and the substrate (right) for three different distances to the substrate. (a)-(b): One layer. (c)-(d): Four layers. (e)-(f): Full disk. Additionally, we show the frequencies corresponding to $\varepsilon_{\rm SiC}=-1$ (red, dashed) and $\varepsilon_{\rm SiC}=-2$ (black, dashed). All parameters as before.}
		\label{fig:tot_spec}
	\end{figure}

As we did before for the two-particle case, we will now discuss the spectra for different configurations: a ring ($N=84$ particles, one layer), a crown ($N=318$, 4 layers), and the full disk ($N=583$). The spectra are shown in Fig.~\ref{fig:tot_spec}. In the left column, panels (a), (c), and (e) show the spectral heat flux between particles and background $P_{{\rm back} \rightarrow {\rm dip}}$ and panels (b), (d), and (f) in the right column the one between the particles and the substrate $P_{{\rm sub} \rightarrow {\rm dip}}$. Everything is shown for the three regimes: near field, far field, and intermediate regime. For $P_{{\rm back} \rightarrow {\rm dip}}$ we find either two [Fig.~\ref{fig:tot_spec}(a), (e)] or three [Fig.~\ref{fig:tot_spec}(c)] major resonance peaks which are highlighted in Fig.~\ref{fig:tot_spec} by the arrows. We can also find the same resonances in $P_{{\rm sub} \rightarrow {\rm dip}}$ together with an additional peak at $\omega_{\rm SPhP}$ (red dashed line) which surpasses the other ones in the near field. Since we can see in each figure in \ref{fig:tot_spec} that this peak dominates in the near field and vanishes in the far field, this proves the strong influence of the coupling between the dipoles of each configuration and the SPhP mode of the substrate. As expected, the eigenmodes of the configuration dominates in the far field for $P_{{\rm sub} \rightarrow {\rm dip}}$ and the overall spectral amplitude goes down as shown in Fig.~\ref{fig:layers}. In the case of $P_{{\rm back} \rightarrow {\rm dip}}$, the spectrum behaves vice versa regarding that distance dependence as explained in the previous Section due to a decreasing attenuation due to coupling with the substrate. Interestingly, the peak at the largest frequency is always more pronounced for the intermediate regime compared to the other two regimes. We also relate this to our observation that the intermediate regime has a different spatial distribution for $P_{{\rm back} \rightarrow {\rm dip}}$. For completeness, we also show by the black dashed line the resonance frequency of a single spherical SiC particle at which $\varepsilon_{\rm SiC}=-2$ holds. It is clear that this frequency does not play a significant role in the overall spectrum.

\subsubsection{Spectrum of a ring with and without substrate} \label{ch:spec_ring}

\begin{figure}[H]
		\centering
		\includegraphics[width=0.5\textwidth]{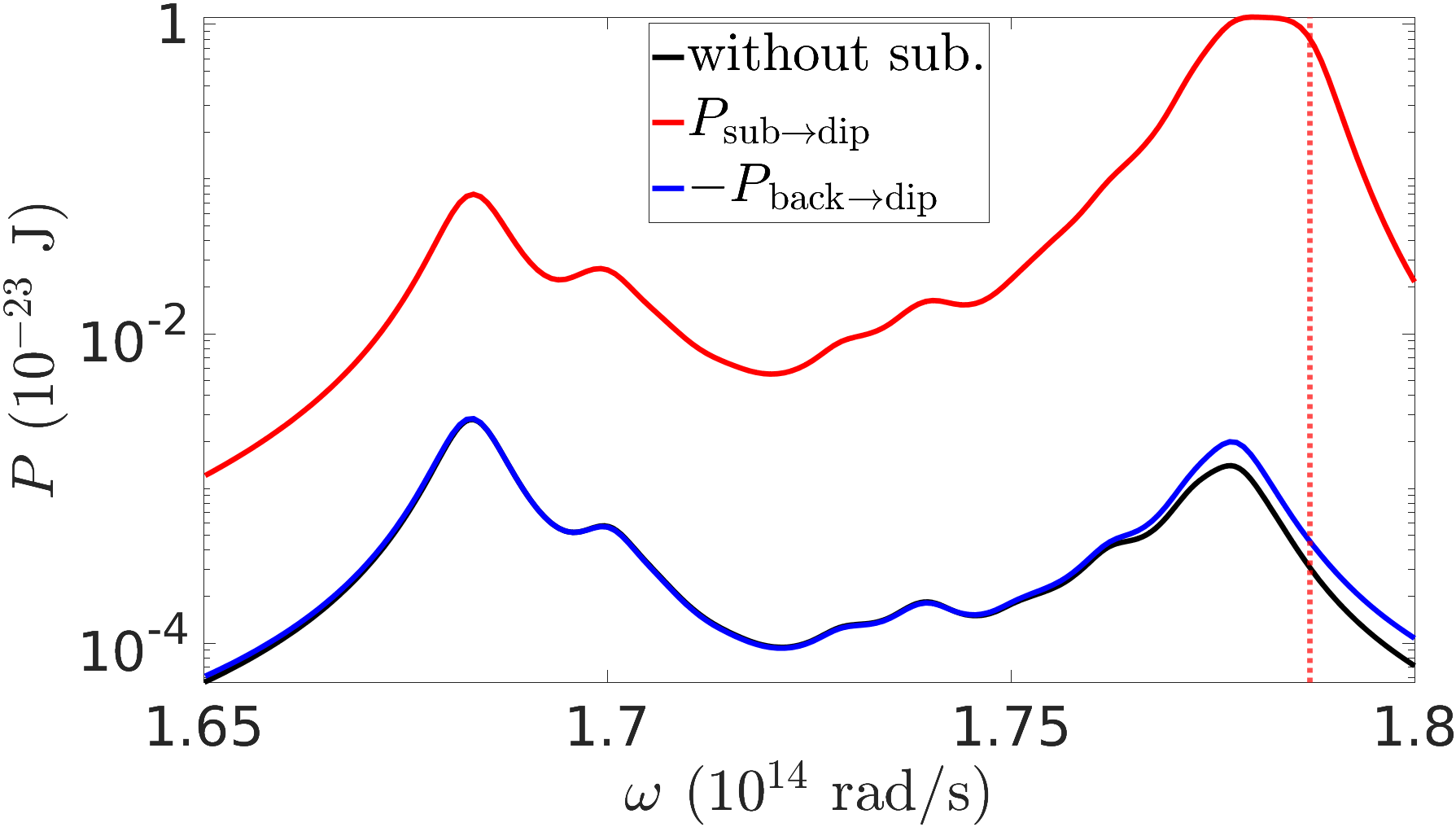}
		\caption{Comparison of the spectral heat fluxes between particles and background (blue) and substrate (red) with the case without substrate (black). The distance between the ring and the substrate is fixed to $d=200$\,nm. All other parameters as before.}
		\label{fig:84_wos_ws}
	\end{figure}

In Fig.~\ref{fig:84_wos_ws} we show the same as in Fig.~\ref{fig:tot_spec}(a)-(b) for $d=200$ nm in comparison with the case without substrate. Let us first of all stress that we can identify all resonances with roots of the determinant in Eq.~\eqref{eq:M}. However, since we have 84 particles for this configuration, we could in principle have 232 different roots in the determinant so that it is not easy to connect each root to one resonance but rather clusters of roots. Comparing all three graphs, one can also clearly see that, apart from the peak highlighted by the red dashed line, all resonances stem from the eigenmodes of the ring since they also appear in the case without substrate. At the red dashed line the SPhP mode has a strong influence on the spectral heat flux between dipoles and substrates. 

Interestingly, $P_{{\rm back} \rightarrow {\rm dip}}$ differs significantly from the case without substrate at the peak at $\omega=1.777 \times 10^{14} \, {\rm rad}/{\rm s}$. We conclude, therefore, that the substrate can enhance this resonance which we also showed in Fig.~\ref{fig:tot_spec}(a) while noticing that this enhancement is particularly important in the intermediate regime of $d = 1\,\mu$m. To go further into details we will now focus on different particles on the ring for further spectral analysis.

\subsubsection{Special particles on a ring} \label{ch:special}

For the ring configuration we compare the spectra of four particles that are highlighted in red in the legend of Fig.~\ref{fig:sp_part}. We compare three different distances, one for each regime, for $P_{{\rm back/sub} \rightarrow {\rm dip}}$ [(a), (c), (e) / (b), (d), (f)]. As a reference, we also put the spectrum for the sum over all particles of the disk in red. The four particles correspond to the one at the top (P$_1$), two ones that are close to the middle (P$_2$ and P$_3$), and one at the very bottom close to the substrate (P$_4$), as already depicted in Fig.~\ref{fig:1l}.

For $P_{{\rm sub} \rightarrow {\rm dip}}$, the red dashed line highlights $\omega_{\rm SPhP}$. At this frequency we clearly observe a strong resonance for the particle close to the substrate (P$_4$) followed by the one in the middle that is closer to the substrate (P$_3$) in the near-field in Fig.~\ref{fig:sp_part}(b). The difference in amplitude is getting less for larger distances and seems to vanish in the far field, which is not surprising in the context of the SPhP mode coupling between substrate and particles. The resonance at $\omega_{1r} = 1.683 \times 10^{14} \, {\rm rad}/{\rm s}$ seems to favor the particles closer to the center. For larger distances this favoring is even more obvious. This explains why we see these particles as bright spots in Fig.~\ref{fig:1l} in the far field. 

For $P_{{\rm back} \rightarrow {\rm dip}}$ we do not observe many differences in the far field; all graphs almost overlap. In the intermediate regime and in the near field, however, the top particles exhibits the strongest resonance at $\omega_{1r}$ and remains dominating up until close to $\omega_{2r}$ which is where the middle particles become more important. Note that at this frequency for the intermediate regime $d=1\,\mu$m, the difference between the centered particles (P$_2$ and P$_3$) and the top particle (P$_1$) increases which coincides with the inversion of graphs for different distances in Fig.~\ref{fig:tot_spec}(a), (c), (e) at this frequency. Finally, we will show the influence of the configurational modes and the SPhP mode spatially.

\begin{widetext}
	
\begin{figure}[H]
		\centering
		\includegraphics[width=0.7\textwidth]{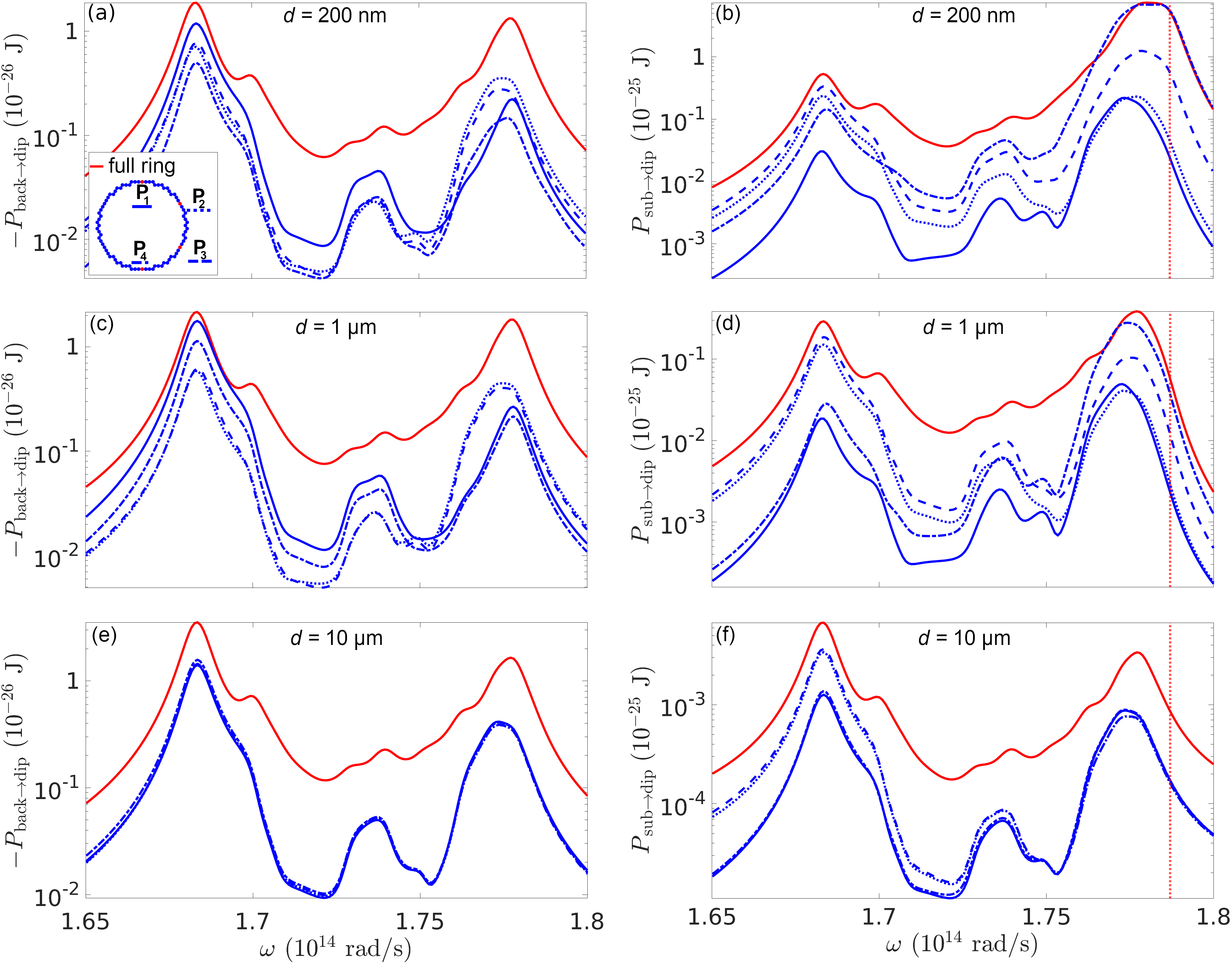}
		\caption{Comparison of the spectral heat fluxes between certain particles (see legend and Fig.~\ref{fig:1l}) and background ((a), (c), and (e)) and substrate ((b), (d), and (f)) for different particles (blue) and the sum of all particles on the ring divided by 15 (red). $\omega_{\rm SPhP}$ is highlighted by the red dashed line at the bottom. All parameters as before.}
		\label{fig:sp_part}
	\end{figure}

\subsubsection{Spatial distribution for special frequencies} \label{ch:spec_all}

Let us finally look at the spatial distribution of the heat fluxes $P_{{\rm back/sub} \rightarrow {\rm dip}}$ for different frequencies corresponding to the major peaks highlighted in Fig.~\ref{fig:tot_spec} for the cases of the ring ($N=84$ particles, one layer), a crown ($N=318$, four layers), and the full disk ($N=583$). The calculations for all figures are performed for $d=200$\,nm. 

For the ring we see the largest heat flux $P_{{\rm back} \rightarrow {\rm dip}}$ at $\omega_{1r}$ at the top particle which is due to the eigenmodes of the ring being less attenuated there. For the other two frequencies we observe the eigenmode signature attenuated for particles closer to the substrate and also the top particles as described in the previous Section. For $\omega_{\rm SPhP}$ and $\omega_{2r}$ the heat flux $P_{{\rm sub} \rightarrow {\rm dip}}$ is the most pronounced at the bottom particle due to strong coupling due to the SPhP mode and evanescent modes in general. At $\omega_{1r}$ we re-encounter the eigenmode signature enhanced for particles closer to the substrate. 
\begin{figure}[H]
		\centering
		\includegraphics[width=0.7\textwidth]{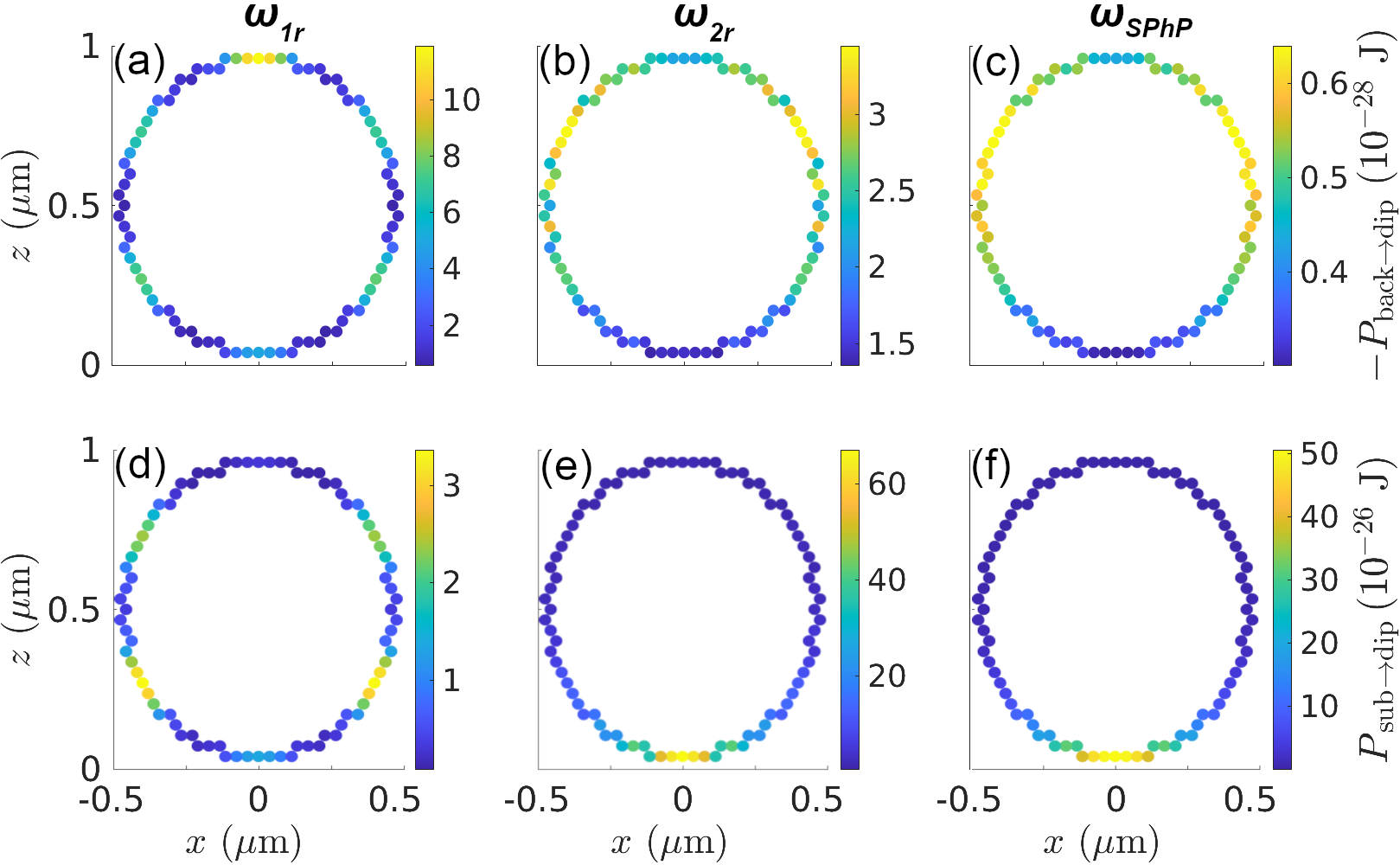}
		\caption{Spatial distribution of the spectral heat fluxes between particles and background ((a)-(c)) and substrate ((d)-(f)) for different frequencies indicated above the upper panel for the ring configuration (one layer). The distance between the ring and the substrate is fixed to $d=200$ nm. The frequencies are the ones highlighted in Fig.~\ref{fig:tot_spec}. All other parameters as before.}
		\label{fig:84_spec}
	\end{figure}

For the cases of a crown and a disk we observe a similar behavior. The SPhP mode has a clear signature by enhancing the $P_{{\rm sub} \rightarrow {\rm dip}}$ for the bottom particles. For the other frequencies we always find the signature of an eigenmode that is enhanced for the bottom particles due to evanescent waves. For $P_{{\rm back} \rightarrow {\rm dip}}$ we also clearly see the eigenmodes of the system but attenuated at the bottom part due to coupling with the substrate. Especially for the full disk the different spatial distribution for each eigenmode is well visible. The only exception is the case of $\omega_{1d}$ for the full disk where we see brighter spots closer to the substrate . 
\begin{figure}[H]
		\centering
		\includegraphics[width=0.93\textwidth]{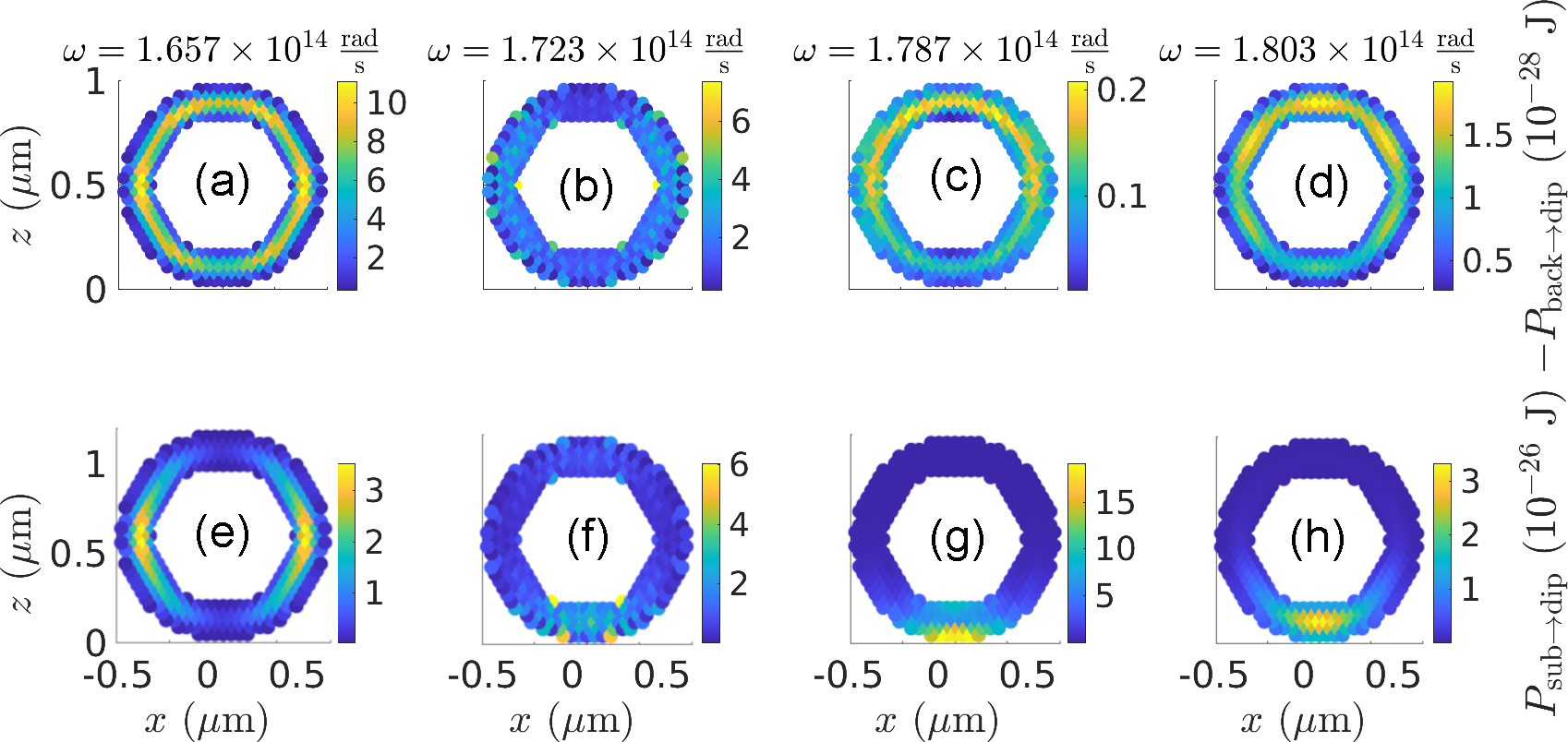}
		\caption{As in Fig.~\ref{fig:84_spec} but for a crown configuration (4 layers).}
		\label{fig:498_spec}
	\end{figure}

Conclusively, this shows the ``competition" between the eigenmodes of the configuration and the coupling with substrate which either enhances the heat flux especially at the SPhP mode or attenuates it for the heat flux between particles and background.
\begin{figure}[H]
		\centering
		\includegraphics[width=0.7\textwidth]{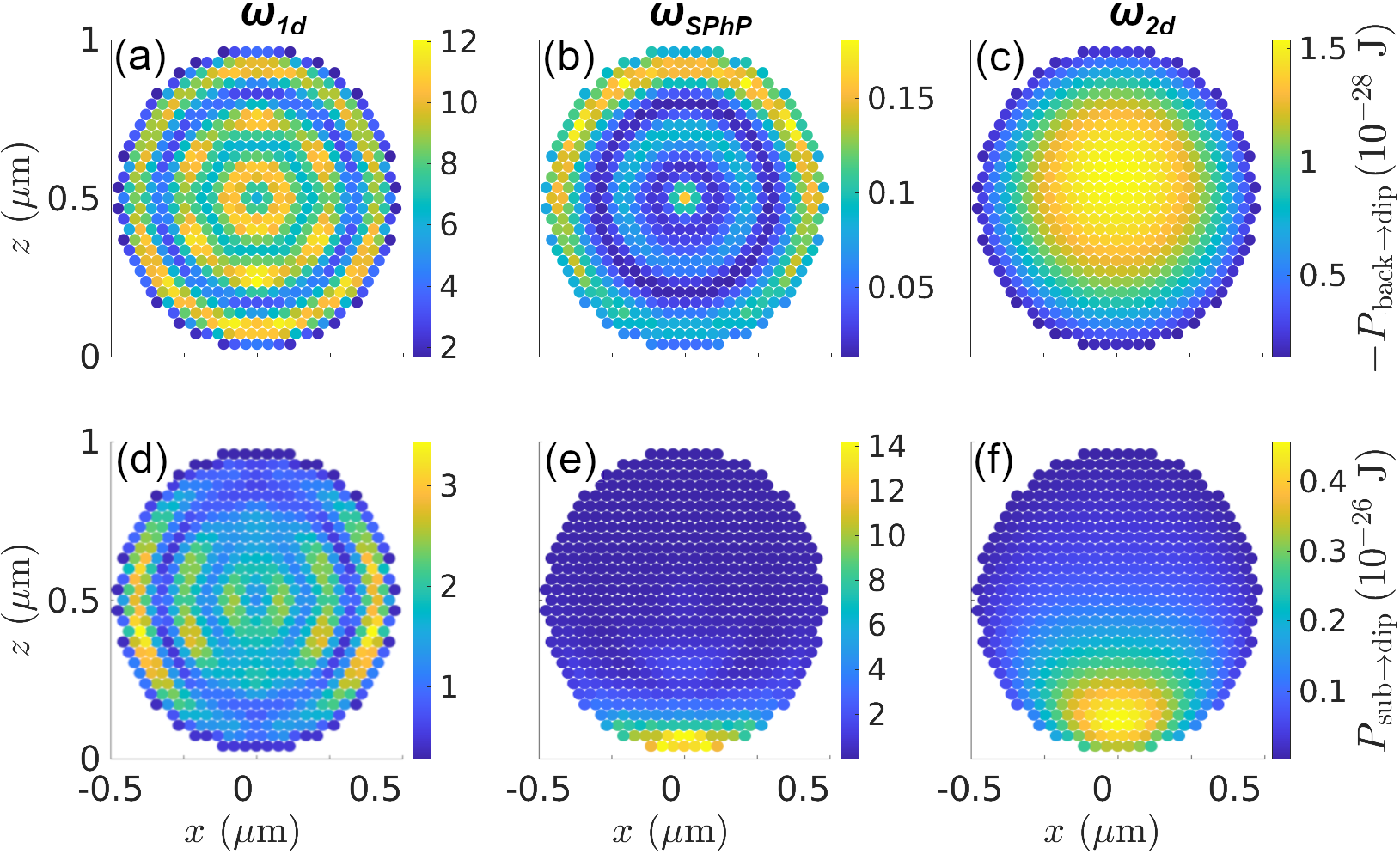}
		\caption{As in Fig.~\ref{fig:84_spec} but for the full disk.}
		\label{fig:583_spec}
	\end{figure}

\end{widetext}

\section{Conclusion} \label{ch:conc}

By performing a tomographic analysis of radiative heat transfers between an object, a substrate, and its thermal environment, we have demonstrated the crucial role played by many-body interactions in near-field regime and highlighted the intimate connection which exists between power distribution and eigenmodes within a solid. Our study sheds light on the fundamental mechanisms driving heat exchanges between mesoscopic objects and their surrounding environment. Our study paves the way to a rational design of local hot spots at a deep sub-wavelength scale by exploiting shape optimization of solids. This could lead to important implications in the fields of nanoscale thermal management, heat-assisted data recording, and nanoscale thermal imaging.

\section{Acknowledgments}

This work was supported by the French Agence Nationale de la Recherche (ANR), under grant ANR-21-CE30-0030 (NBODHEAT). Additionally, F. H. acknowledges financial support by the Walter Benjamin Program of the Deutsche Forschungsgemeinschaft (eng. German Research Foundation) under project number 519479175.

\appendix

\section{General Green's functions and correlation functions}\label{AppA}

As mentioned in section \ref{ch:Th_fram}, the Green's function for each particle consists of a vacuum part and a reflected part due to the presence of the substrate~\cite{Sipe_1987}. In Eq.~\eqref{eq:Green} this is described more explicitly. Throughout this work, we use the vacuum Green's function
\begin{align}
\mathds{G}_{{\rm E}, 0} (\mathbf{r}, \mathbf{r}') & = \frac{e^{{\rm i} k_0 \rho}}{4 \pi \rho} \biggl[ \frac{k_0^2 \rho^2 + {\rm i} k_0 \rho - 1}{k_0^2 \rho^2} \mathds{1} - \frac{k_0^2 \rho^2 + 3 {\rm i} k_0 \rho - 3}{k_0^2 \rho^2} \hat{\boldsymbol{\rho}} \otimes \hat{\boldsymbol{\rho}} \biggr]
\end{align}
with 
\begin{align}
\hat{\boldsymbol{\rho}} & = \frac{\mathbf{r} - \mathbf{r}'}{|\mathbf{r} - \mathbf{r}'|}, \\
\rho & = |\mathbf{r} - \mathbf{r}'|.
\end{align}
For the reflected contribution, we employ the following expression
\begin{align}
\mathds{G}_{\rm E, s} (\mathbf{r}, \mathbf{r}') & = \int \frac{{\rm d}^2 k_\perp}{(2 \pi)^2} e^{{\rm i} \mathbf{k}_\perp \cdot (\mathbf{x} - \mathbf{x}')} \frac{{\rm i} e^{{\rm i} k_z (z + z')}}{2 k_z} \sum_{n = s,p} r_i \mathbf{a}_i^{+} \otimes \mathbf{a}_i^{-} 
\label{eq:Gs}
\end{align}
with the polarization unit vectors
\begin{align}
\mathbf{a}_s^{\pm} & = \frac{1}{k_\perp} (-k_y, k_x, 0)^t, \\
\mathbf{a}_p^{\pm} & = \frac{1}{k_\perp k_0} (\pm k_x k_z, \pm k_y k_z, -k_\perp^2)^t
\end{align}
and the Fresnel amplitude reflection coefficients
\begin{align}
r_s & = \frac{k_z - k_{z, {\rm sub}}}{k_z + k_{z, {\rm sub}}}, \\
r_p & = \frac{\varepsilon_{\rm sub} k_z - k_{z, {\rm sub}}}{\varepsilon_{\rm sub} k_z + k_{z, {\rm sub}}}.
\label{eq:rp}
\end{align}
Therein, we used the definitions
\begin{align}
\mathbf{k}_\perp & = (k_x, k_y)^t, \\
\mathbf{x} & = (x, y)^t, \\
{\rm d}^2 k_\perp & = {\rm d} k_x {\rm d} k_y, \\
k_z & = \sqrt{k_0^2 - k_\perp^2}, \\
k_{z, {\rm sub}} & = \sqrt{\varepsilon_{\rm sub} k_0^2 - k_\perp^2}.
\end{align}

\section{Green's function and correlation functions for two particles above a substrate}\label{AppB}

The reflected part of the Green's function in Eq.~\eqref{eq:Gs} simplifies for two particles as in Fig.~\ref{fig:2p} to
\begin{align}
\mathds{G}_{\rm E, s} (\mathbf{r}_{1/2}, \mathbf{r}_{1/2}) & = G_{{\rm E,s},\perp}^{\rm id} \left( \hat{\mathbf{x}} \otimes \hat{\mathbf{x}} + \hat{\mathbf{y}} \otimes \hat{\mathbf{y}} \right) + G_{\rm E,s,z}^{\rm id} \hat{\mathbf{z}} \otimes \hat{\mathbf{z}}
\label{eq:Gs_2ws}
\end{align}
with 
\begin{align}
G_{{\rm E,s},\perp}^{\rm id} & = {\rm i} \int_0^\infty \frac{{\rm d} k_\perp k_\perp}{8 \pi k_z} e^{2 {\rm i} k_z (d+R)} \biggl[ r_s - r_p \frac{k_z^2}{k_0^2} \biggr] , \\
G_{\rm E,s,z}^{\rm id} & = {\rm i} \int_0^\infty \frac{{\rm d} k_\perp k_\perp^3}{4 \pi k_0^2 k_z} e^{2 {\rm i} k_z (d+R)} r_p
\end{align}
if both spacial arguments of the Green's function are identical and to
\begin{align}
\mathds{G}_{\rm E, s} (\mathbf{r}_{1/2}, \mathbf{r}_{2/1}) & = G_{\rm E,s,x}^{\rm dif} \hat{\mathbf{x}} \otimes \hat{\mathbf{x}} + G_{\rm E,s,y}^{\rm dif} \hat{\mathbf{y}} \otimes \hat{\mathbf{y}} + G_{\rm E,s,z}^{\rm dif} \hat{\mathbf{z}} \otimes \hat{\mathbf{z}} \notag \\
& \quad \pm G_{\rm E,s,mix} \left[ \hat{\mathbf{x}} \otimes \hat{\mathbf{z}} - \hat{\mathbf{z}} \otimes \hat{\mathbf{x}} \right] 
\end{align}
with 
\begin{align}
G_{\rm E,s,x/y}^{\rm dif} & = {\rm i} \int_0^\infty \frac{{\rm d} k_\perp k_\perp}{8 \pi k_z} e^{2 {\rm i} k_z (d+R)} \biggl[ r_s \left(J_0(k_\perp l) \pm J_2(k_\perp l) \right) \notag \\
& \quad - r_p \frac{k_z^2}{k_0^2} \left(J_0(k_\perp l) \mp J_2(k_\perp l) \right) \biggr] 
\end{align}
as well as
\begin{align}
G_{\rm E,s,z}^{\rm dif} & = {\rm i} \int_0^\infty \frac{{\rm d} k_\perp k_\perp^3}{4 \pi k_0^2 k_z} e^{2 {\rm i} k_z (d+R)} r_p J_0(k_\perp l) , \\
G_{\rm E,s,mix} & = \int_0^\infty \frac{{\rm d} k_\perp k_\perp^2}{4 \pi k_0^2} e^{2 {\rm i} k_z (d+R)} r_p J_1(k_\perp l)
\end{align}
otherwise. For the correlation function of the background fields, we obtain
\begin{align}
\mathds{C}_{\text{b},11/22} & = \mathcal{C}_{{\rm b},\perp}^{\rm id} \left( \hat{\mathbf{x}} \otimes \hat{\mathbf{x}} + \hat{\mathbf{y}} \otimes \hat{\mathbf{y}} \right) + \mathcal{C}_{\rm b,z}^{\rm id} \hat{\mathbf{z}} \otimes \hat{\mathbf{z}}
\end{align}
with 
\begin{align}
C_{{\rm b},\perp}^{\rm id} & = \int_0^{k_0} \frac{\text{d} k_\perp k_\perp}{8 \pi k_z} \biggl[ R_s^{+} (k_\perp, d) + R_p^{-} (k_\perp, d) \frac{k_z^2}{k_0^2} \biggr] , \\
C_{\rm b,z}^{\rm id} & = \int_0^{k_0} \frac{\text{d} k_\perp k_\perp^3}{4 \pi k_z k_0^2} R_p^{+} (k_\perp, d)
\end{align}
in the case of identical particles and 
\begin{align}
\mathds{C}_{\text{b},12/21} & = \mathcal{C}_{\rm b,x}^{\rm dif} \hat{\mathbf{x}} \otimes \hat{\mathbf{x}} + \mathcal{C}_{\rm b,y}^{12/21} \hat{\mathbf{y}} \otimes \hat{\mathbf{y}} + \mathcal{C}_{\rm b,z}^{\rm dif} \hat{\mathbf{z}} \otimes \hat{\mathbf{z}} \notag \\
& \quad + \mathcal{C}_{\rm b,xz}^{12/21} \hat{\mathbf{x}} \otimes \hat{\mathbf{z}} + \mathcal{C}_{\rm b,zx}^{12/21} \hat{\mathbf{z}} \otimes \hat{\mathbf{x}} 
\end{align}
with 
\begin{align}
C_{\rm b,x/y}^{\rm dif} & = \int_0^{k_0} \frac{\text{d} k_\perp k_\perp}{8 \pi k_z} \biggl[ R_s^{+} (k_\perp, d) \left[ J_0(k_\perp l) \pm J_2(k_\perp l) \right] \notag \\
& \quad + R_p^{-} (k_\perp, d) \frac{k_z^2}{k_0^2} \left[ J_0(k_\perp l) \mp J_2(k_\perp l) \right] \biggr] , \\
C_{\rm b,z}^{\rm dif} & = \int_0^{k_0} \frac{\text{d} k_\perp k_\perp^3}{4 \pi k_z k_0^2} R_p^{+} (k_\perp, d) J_0 (k_\perp l)
\end{align}
as well as
\begin{align}
C_{\rm b,xz}^{12/21} & = \pm \text{i} \int_0^{k_0} \frac{\text{d} k_\perp k_\perp^2}{4 \pi k_0^2} \tilde{R}_p^{-} (k_\perp, d) J_1 (k_\perp l) , \\
C_{\rm b,zx}^{12/21} & = \pm \text{i} \int_0^{k_0} \frac{\text{d} k_\perp k_\perp^2}{4 \pi k_0^2} \tilde{R}_p^{+} (k_\perp, d) J_1 (k_\perp l) 
\end{align}
if one considers different particles. Here, we introduced
\begin{align}
R_i^{\pm} (k_\perp, d) & = 1 + |r_i|^2 \pm 2 \text{Re} \left(e^{2 \text{i} k_z (d+R)} r_i \right) , \\
\tilde{R}_i^{\pm} (k_\perp, d) & = 1 - |r_i|^2 \pm 2 \text{i} \text{Im} \left(2 e^{\text{i} k_z (d+R)} r_i \right) .
\end{align}
In the same way we obtain for the correlation function of the substrate fields
\begin{align}
\mathds{C}_{{\rm s},11/22} & = \mathcal{C}_{{\rm s},\perp}^{\rm id} \left( \hat{\mathbf{x}} \otimes \hat{\mathbf{x}} + \hat{\mathbf{y}} \otimes \hat{\mathbf{y}} \right) + \mathcal{C}_{\rm s,z}^{\rm id} \hat{\mathbf{z}} \otimes \hat{\mathbf{z}} \\
\end{align}
with 
\begin{align}
C_{{\rm s},\perp}^{\rm id} & = \int_0^{k_0} \frac{\text{d} k_\perp k_\perp}{8 \pi k_z} \biggl[ 1 - |r_s|^2 + \frac{k_z^2}{k_0^2} (1 - |r_p|^2) \biggr] \notag \\
& \quad + \int_{k_0}^\infty \frac{\text{d} k_\perp k_\perp}{4 \pi |k_z|} e^{- 2 |k_z| (d+R)} \biggl[ \text{Im}(r_s) + \frac{|k_z|^2}{k_0^2} \text{Im}(r_p) \biggr] 
\end{align}
and
\begin{align}
C_{\rm s,z}^{\rm id} & = \int_0^{k_0} \frac{\text{d} k_\perp k_\perp^3}{4 \pi k_z k_0^2} (1 - |r_p|^2) \notag \\
& \quad + \int_{k_0}^\infty \frac{\text{d} k_\perp k_\perp^3}{2 \pi |k_z| k_0^2} e^{- 2 |k_z| d} \text{Im}(r_p) 
\end{align}
for the case of two identical particles and 
\begin{align}
\mathds{C}_{\text{s},12/21} & = \mathcal{C}_{\rm s,x}^{\rm dif} \hat{\mathbf{x}} \otimes \hat{\mathbf{x}} + \mathcal{C}_{\rm s,y}^{\rm dif} \hat{\mathbf{y}} \otimes \hat{\mathbf{y}} + \mathcal{C}_{\rm s,z}^{\rm dif} \hat{\mathbf{z}} \otimes \hat{\mathbf{z}} \notag \\
& \quad + \mathcal{C}_{\rm s,xz}^{12/21} \hat{\mathbf{x}} \otimes \hat{\mathbf{z}} + \mathcal{C}_{\rm s,zx}^{12/21} \hat{\mathbf{z}} \otimes \hat{\mathbf{x}}
\end{align}
with
\begin{align}
C_{\rm s,x/y}^{\rm dif} & = \int_0^{k_0} \frac{\text{d} k_\perp k_\perp}{8 \pi k_z} \biggl[ \Gamma_{+} J_0(k_\perp l) \pm \Gamma_{-} J_2(k_\perp l) \biggr] \notag \\
& \quad + \int_{k_0}^\infty \frac{\text{d} k_\perp k_\perp}{4 \pi |k_z|} e^{- 2 |k_z| (d+R)} \biggl[ \Delta_{+} J_0(k_\perp l) \pm \Delta_{-} J_2(k_\perp l) \biggr] \\
C_{\rm s,z}^{\rm dif} & = \int_0^{k_0} \frac{\text{d} k_\perp k_\perp^3}{4 \pi k_z k_0^2} (1 - |r_p|^2) J_0 (k_\perp l) \notag \\
& \quad + \int_{k_0}^\infty \frac{\text{d} k_\perp k_\perp^3}{2 \pi |k_z| k_0^2} e^{- 2 |k_z| (d+R)} \text{Im}(r_p) J_0 (k_\perp l)
\end{align}
and
\begin{align}
C_{\rm s,xz}^{12/21} & = \mp \text{i} \int_0^{k_0} \frac{\text{d} k_\perp k_\perp^2}{4 \pi k_0^2} (1 - |r_p|^2) J_1 (k_\perp l) \notag \\
& \quad \pm \int_{k_0}^\infty \frac{\text{d} k_\perp k_\perp^2}{2 \pi k_0^2} e^{- 2 |k_z| (d+R)} \text{Im}(r_p) J_1 (k_\perp l) , \\
C_{\rm s,zx}^{12/21} & = \mp \text{i} \int_0^{k_0} \frac{\text{d} k_\perp k_\perp^2}{4 \pi k_0^2} (1 - |r_p|^2) J_1 (k_\perp l) \notag \\
& \quad \mp \int_{k_0}^\infty \frac{\text{d} k_\perp k_\perp^2}{2 \pi k_0^2} e^{- 2 |k_z| (d+R)} \text{Im}(r_p) J_1 (k_\perp l)
\end{align}
in the case of two different particles. Here, we introduced 
\begin{align}
\Gamma_{\pm} & = 1 - |r_s|^2 \pm \frac{k_z^2}{k_0^2} (1 - |r_p|^2) , \\
\Delta_{\pm} & = \text{Im}(r_s) \pm \frac{|k_z|^2}{k_0^2} \text{Im}(r_p) .
\label{eq:Dpm}
\end{align}

\bibliography{Referenzen}

\end{document}